\documentclass[12pt]{article}
\pdfoutput=1
\usepackage{amssymb}
\usepackage{amsmath}
\usepackage{graphicx}
\usepackage{booktabs}
\usepackage{color}
\usepackage{textcomp}
\usepackage{multirow}
\usepackage{bm}
\usepackage{caption}
\usepackage{subcaption}
\usepackage{tikz}
\usepackage{longtable}
\usepackage{colortbl}

\usepackage{jheppub}

\newcommand{\eq}[1]{equation~(\ref{#1})}
\newcommand{\fig}[1]{Figure~\ref{#1}}
\newcommand{\tab}[1]{Table~\ref{#1}}
\newcommand{\sect}[1]{Section~\ref{#1}}
\newcommand{\tr}{\mathrm{tr}\,}

\begin{document}
\begin{flushright}
\small{
WUB/15-04
}
\end{flushright}

\title{Five-Dimensional Gauge-Higgs Unification:\\ A Standard Model-Like Spectrum}

\author[a]{Maurizio~Alberti,} \emailAdd{alberti@uni-wuppertal.de}
\author[b]{Nikos~Irges,} \emailAdd{irges@mail.ntua.gr}
\author[a]{Francesco~Knechtli,} \emailAdd{knechtli@physik.uni-wuppertal.de}
\author[a]{Graham~Moir} \emailAdd{gmoir@uni-wuppertal.de}

\affiliation[a]{Department of Physics, Bergische Universit\"at Wuppertal\\Gaussstr. 20, D-42119 Wuppertal, Germany}
\affiliation[b]{Department of Physics, National Technical University of Athens\\Zografou Campus, GR-15780 Athens, Greece}

\abstract{We study the viability of five-dimensional gauge theories as candidates for the origin of the
Higgs field and its mechanism for spontaneous symmetry breaking. Within the framework of lattice
field theory, we consider the simplest model of an $SU(2)$ gauge theory. We construct this theory on
a five-dimensional orbifold which explicitly breaks the gauge symmetry to $U(1)$ at the fixed points of the orbifold.
Using anisotropic gauge couplings, we find that this theory exhibits three distinct phases which we label as confined, Higgs and hybrid.
Within the Higgs phase, close to the Higgs-hybrid phase transition, we find that the ratio of the Higgs to gauge boson masses
takes Standard Model-like values. Precisely in this region of the phase diagram, we find dimensional reduction via localisation.}


\maketitle

\section{Introduction \label{s_intro}}

The Higgs sector of the Standard Model is an effective theory that seems to be consistent with observations, 
but at the same time hides a number of puzzles. One such puzzle is the origin of the potential responsible for
Spontaneous Symmetry Breaking (SSB), which leads to the Brout-Englert-Higgs (BEH) mechanism \cite{Englert:1964et,Higgs:1964ia,Higgs:1964pj}.
Another is the mechanism behind the stability of the Higgs mass under quantum fluctuations; its
quadratic sensitivity to an ultra-violet cut-off is related to the so-called hierarchy problem. A class of extensions
to the Standard Model aimed at addressing these puzzles by the use of extra dimensions come under the heading of
Gauge-Higgs Unification (GHU) \cite{Manton:1979kb,Fairlie:1979at,Hosotani:1983vn}. In these models, the Higgs
field originates from extra-dimensional components of the gauge field and gives rise to massive gauge bosons in the regular
four dimensions. In this study, we consider the simplest case of one extra dimension where, due to the higher dimensional gauge invariance,
the Higgs potential remains zero at tree level and is generated only through quantum effects \cite{Coleman:1973jx}. Furthermore, the use
of an orbifold geometry \cite{Hebecker:2001jb} can break the gauge group at the fixed points, allowing for a Standard Model-like Higgs field
in the fundamental representation. 

In perturbation theory, where the extra dimension is compactified and the five-dimensional fields are expanded as a
series of four-dimensional Kaluza-Klein modes, the Higgs mass is independent of the cut-off (at one-loop) \cite{vonGersdorff:2002as,Cheng:2002iz}. However,
we consider a pure gauge theory where it is known that the one-loop effective Higgs potential \cite{Antoniadis:2001cv} does not develop a symmetry
breaking minimum \cite{Kubo:2001zc}. Although the Higgs mass typically tends to be light \cite{Scrucca:2003ra}, one way to circumvent this problem
is to include fermionic degrees of freedom which induce SSB via the Hosotani mechanism \cite{Hosotani:1983vn,Cossu:2013ora}. Alternatively, by introducing
an explicit cut-off in pure gauge theory, a one-loop potential with a symmetry breaking minimum can indeed be generated \cite{Irges:2007qq}; in the presence
of this cut-off, the perturbative non-renormalisability of the five-dimensional theory can not be ignored, and the stability of the Higgs potential beyond
one-loop has to be investigated. \\

In the context of lattice field theory, five-dimensional GHU models were formulated in \cite{Irges:2004gy,Knechtli:2005dw} for an orbifold geometry.
Subsequently, non-perturbative studies using Monte Carlo simulations of the path integral showed that SSB occurs in pure gauge 
theory \cite{Irges:2006zf,Irges:2006hg}. This observation was confirmed in \cite{Irges:2012ih,Irges:2012mp} via semi-analytic
mean-field calculations \cite{Drouffe:1983fv}. An immediate concern here is the apparent disagreement between these results and Elitzur's
theorem \cite{Elitzur:1975im}, which states that a local gauge symmetry can not be spontaneously broken. However, their reconciliation
was confirmed by the discovery of a global symmetry in \cite{Ishiyama:2009bk}; it is the spontaneous breaking of this global symmetry
which leads to the BEH mechanism in the context of non-perturbative GHU models \cite{Irges:2013rya}. Other previous explorations of this theory have
focused on the case of a toroidal geometry which is known to posses confined and de-confined phases separated by a first-order phase
transition \cite{Creutz:1979dw,Knechtli:2011gq,Farakos:2010ie,DelDebbio:2013rka,Irges:2015uta}; 
second-order transitions due to compactification were studied in \cite{Ejiri:2000fc,deForcrand:2010be,Knechtli:2011gq}, and the scalar spectrum was measured
in \cite{DelDebbio:2012mr}.

In this paper we present the phase diagram and spectrum of the theory formulated using an orbifold geometry and an anisotropic lattice. We concentrate on the region where the
the lattice spacing in the usual four dimensions, $a_{4}$, is less than or equal to the lattice spacing in the extra dimension, $a_{5}$. Within this region, the theory exhibits
three distinct phases, which we label as confined, Higgs and hybrid. We find a layered-like dimensional reduction
in the hybrid phase, while in the Higgs phase, close to the Higgs-hybrid phase transition, we observe dimensional reduction on the boundaries via localisation. Significantly, within this region 
close to the Higgs-hybrid phase transition, we find that the spectrum resembles that of the Higgs sector. This theory is therefore a candidate to reproduce the Higgs
sector of the Standard Model.

\section{Theoretical Framework \label{sec:definitions}}

This study will focus on a pure $SU(2)$ Yang-Mills theory in five dimensions. 
We label our continuum five-potential $A_{M}$, where $M = 0, 1, 2, 3, 5,$ corresponding
to the usual four-dimensions and an extra fifth dimension. We construct the theory using an  
orbifold geometry following \cite{Irges:2004gy, Irges:2006hg}, where it was found that this
theory exhibits SSB.

\subsection{Gauge Theory on the Lattice Orbifold \label{s_latorb}}

In order to construct the theory on an orbifold, one first considers a gauge theory
constructed on a five-dimensional periodic Euclidean lattice $T \times L^{3} \times 2N_{5}$, where
$T$ refers to the number of lattice points in the temporal direction, $L$ to the number of lattice points in the spatial
directions and $2N_{5}$ to the number of lattice points along the extra dimension. We denote the lattice spacing as $a$ and
label the coordinates of the lattice points via a set of integers $n \equiv \{n_{M}\}$. In what follows, $\mu$ is used to index the standard four dimensions.
The gauge field is defined as the set of parallel transporters (gauge links) $\{U(n,M) \in SU(2)\}$, identified by lattice
coordinates and direction, that connect neighbouring lattice points.
The orbifolding of the extra dimension is then achieved by a combination of a reflection $\mathcal{R}$ and group conjugation $\mathcal{T}_{g}$,
hence leading us to the orbifolding condition
\begin{equation}\label{eqn:orbifold_condition}
(1 - \mathcal{RT}_{g}) U(n,M) = 0~,
\end{equation}
which is essentially a $\mathbb{Z}_{2}$ projection of the gauge links. The reflection operation acts on both
the lattice points
\begin{equation}\label{eq:refl1}
\bar{n} \equiv \mathcal{R}n = (n_{\mu}, -n_{5})
\end{equation}
and the gauge links
\begin{equation}\label{eq:refl2}
\mathcal{R}U(n,\mu) = U(\bar{n}, \mu)\,,  ~~~~~ \mathcal{R}U(n, 5) = U^{\dagger}(\bar{n} - \hat{5}, 5)~.
\end{equation}
The group conjugation acts solely on the gauge links
\begin{equation}
\mathcal{T}_{g}U(n,M) = g~U(n,M)~g^{-1}~,
\end{equation}
where $g$ is a constant $SU(2)$ matrix such that $g^{2}$ is an element of the centre of $SU(2)$. Without loss of generality, 
we \textit{choose} $g = -i\sigma^{3}$, and in accordance with gauge invariance the gauge group is broken down to $U(1)$ at the
fixed points of the orbifold ($n_{5} = 0, N_{5}$).

\begin{figure}[t!]
  \begin{center}
   \includegraphics[width=0.7\textwidth]{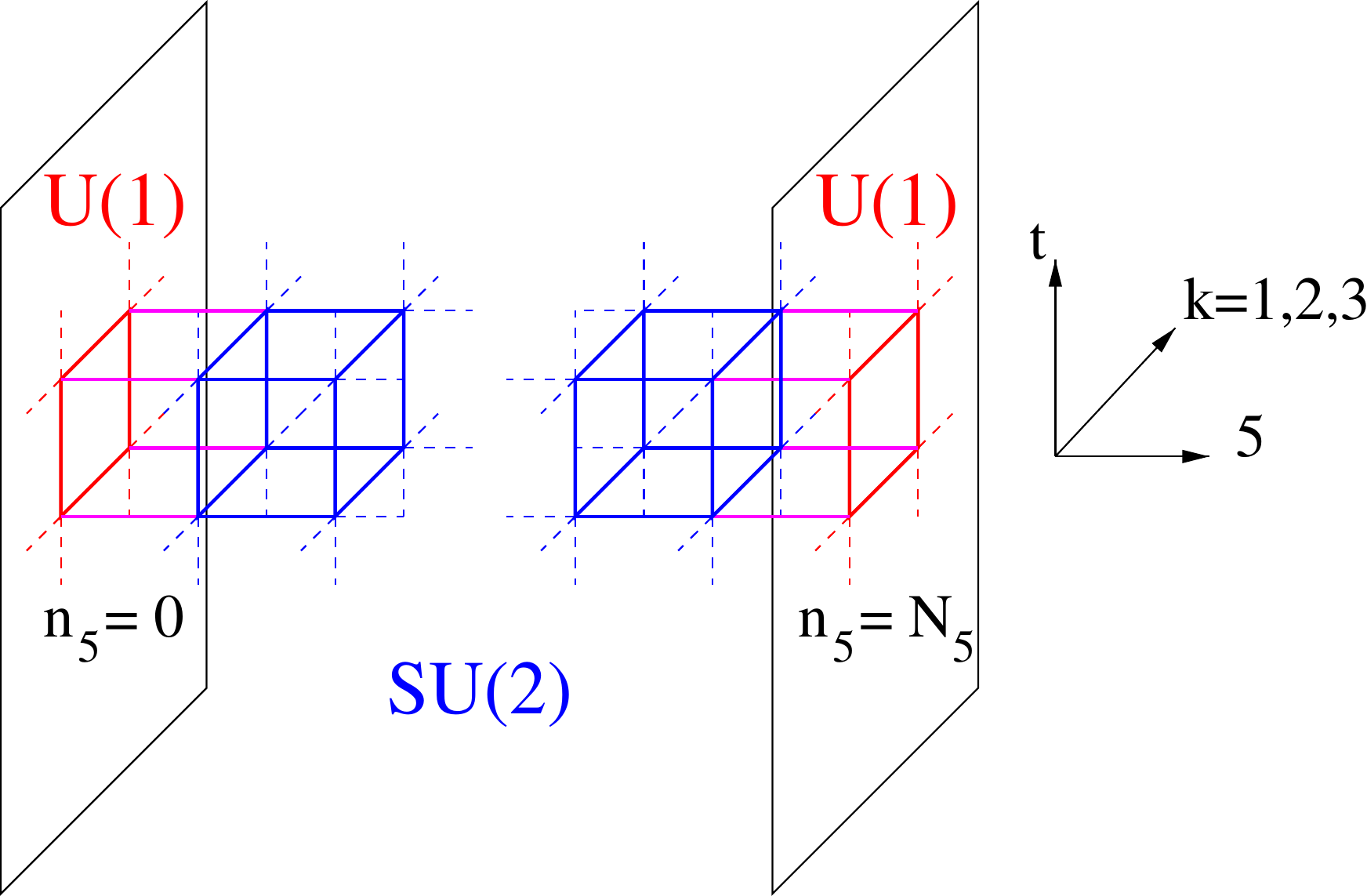}
  \end{center}
  \caption{\footnotesize A sketch of the orbifold lattice and the gauge
links: boundary $U(1)$ links are red, hybrid $SU(2)$ links
sticking to the boundaries are magenta and bulk $SU(2)$ links are blue.
    }
  \label{fig:latorb}
\end{figure}

The theory is now defined in the domain $I = \{ n_{\mu}, 0 \leq n_{5} \leq N_{5} \}$ with volume $T \times L^{3} \times N_{5}$
corresponding to \fig{fig:latorb}. We perform our study using a five-dimensional anisotropic Wilson action
\begin{equation}\label{eqn:wilson_action}
S_W^{orb} = \frac{\beta}{2} \sum_{n_{\mu}} {\left[ \frac{1}{\gamma}\sum_{\mu<\nu}{w~\tr{\left\{1 - U_{\mu\nu}(n_{\mu}) \right\}}} + \gamma \sum_\mu{\tr{\left\{ 1 - U_{\mu5}(n_{\mu}) \right\}}} \right]}~,
\end{equation}
where $w = 1/2$ for plaquettes, $U_{\mu \nu}$, living at the fixed points of the orbifold and $w = 1$ otherwise \cite{Irges:2004gy}. The anisotropy parameter allows for 
different lattice spacings $a_{4}$ and $a_{5}$, where $a_{4}$ denotes the lattice spacing in the usual four dimensions and $a_{5}$ denotes the lattice spacing
in the extra dimension. In the classical limit $\gamma = a_{4}/a_{5}$ and $\beta = 4a_{4} / g^{2}_{5}$, where $g_{5}$ is the dimensionful continuum gauge coupling.
In what follows, we will use an equivalent pair of couplings $\beta_{4}$ and $\beta_{5}$ which are related to the couplings in \eq{eqn:wilson_action} via
\begin{equation}\label{eqn:couplings}
\beta_{4} = \frac{\beta}{\gamma} ~~~~~;~~~~~ \beta_{5} = \beta \gamma ~ .
\end{equation}
As depicted in \fig{fig:latorb}, there are three types of gauge links: four-dimensional $U(1)$ links residing at the fixed points of the orbifold, bulk $SU(2)$ links
and extra-dimensional links which have one end at a fixed point and the other in the bulk; these so-called \textit{hybrid links}
gauge transform as $U \rightarrow \Omega^{U(1)} U \Omega^{\dag SU(2)}$ at the left boundary ($n_{5} = 0$) and $U \rightarrow \Omega^{SU(2)} U \Omega^{\dag U(1)}$
at the right boundary ($n_{5} = N_{5}$).

We note that the orbifold boundary conditions for the gauge field, when the matrix $g$ is chosen to be
the identity, are equivalent to open boundary conditions in the orbifolded direction, cf. \cite{Luscher:2011kk}.

\subsection{Spontaneous Symmetry Breaking on the Orbifold}\label{sec:ssb}

Global symmetries play a crucial role in a non-perturbative gauge invariant formulation of gauge theories.
Elitzur's theorem dictates that any physical effect associated with the breaking of
a local symmetry must originate from the spontaneous breaking of a non-trivial global symmetry \cite{Elitzur:1975im}.

Within our set-up, the spontaneous breaking of the boundary $U(1)$ gauge
symmetry is governed by a so-called
\textit{stick symmetry}, $\mathcal{S}$ \cite{Ishiyama:2009bk}. This symmetry
is defined by the combination
$\mathcal{S} = \mathcal{S}_{L} \cdot \mathcal{S}_{R}$, where
$\mathcal{S}_{L}$ is a symmetry defined on the left boundary via
\begin{eqnarray}
&& U(n_{5} = 0, 5)   \longrightarrow g_{s}^{-1}~U(n_{5} = 0, 5) \label{eq:st1}\\
&& U(n_{5} = 0, \mu) \longrightarrow g_{s}^{-1}~U(n_{5} = 0, \mu)~g_{s}~. \label{eq:st2}
\end{eqnarray}
The symmetry $\mathcal{S}_{L}$, which is not a gauge or centre transformation, is generated by an element $g_{s}$ of the generalized
Weyl group, $W_{SU(2)}(U(1)) = N_{SU(2)}(U(1))/U(1)$, which is the coset of the normaliser of $U(1)$ in $SU(2)$ divided by $U(1)$ \cite{Irges:2013rya}.
It is a $\mathbb{Z}_2$ symmetry (see Appendix \ref{app_stick} for a proof) obeying $\{g, g_{s}\}=0$ that acts only on hybrid and boundary links
\footnote{When $g$ is the identity there is no stick symmetry since it always commutes with $g_s$. The stick symmetry appears only when the scalar left on the boundary by the
orbifold projection is not in the adjoint representation.}. $\mathcal{S}_{R}$ can be defined on the right boundary in an equivalent fashion.

By construction, the system has an additional symmetry, $\mathcal{F}$,
defined as the reflection about the centre of the extra dimension. It is
important to note that $\mathcal{S}$ commutes with $\mathcal{F}$, and we
only consider operators that respect $\mathcal{F}$.
Since it is odd under $\mathcal{S}$, the order parameter for the
spontaneous breaking of the $U(1)$ is
\begin{equation}\label{eq:FZ}
\mathcal{Z}^{\mathcal{F}}_{k} =  \mathcal{Z}_{k} + \mathcal{F}[ \mathcal{Z}_{k} ]~,
\end{equation}
where $\mathcal{Z}_{k}$ is defined via \eq{eqn:Z}.
A non-trivial expectation value of $\mathcal{Z}^{F}_{k}$
signals SSB giving rise to a mass of the associated gauge boson.

An important property of the orbifold, as shown in Appendix \ref{app_stick}, is that the transformation by the centre
of $SU(2)$, which is obtained by applying equations \eqref{eq:st1} and \eqref{eq:st2} twice, does not seem to carry any
physical consequence. All operators associated with bosonic degrees of freedom that extend into the extra dimension are 
even under the centre symmetry, as they must contain an even number of hybrid links. Since there is no order parameter
which is odd under this symmetry, finite temperature phase transitions (defined by the spontaneous breaking of the centre)
cannot occur along the extra dimension of the orbifold. The direct consequence is that the masses associated with the 
orbifold operators can \textit{not} be finite temperature (or Debye \cite{Arnold:1995bh}) masses.

Following the arguments presented in Appendix \ref{app_stick}, the stick symmetry appears to be valid only at finite lattice spacing.
In the perturbative continuum limit \footnote{According to our results, this is the only continuum limit which
exists, since the phase transition is first-order everywhere.} the symmetry, and hence SSB, disappears. This explains the absence of SSB in
pure gauge perturbative models of GHU \cite{Kubo:2001zc}. As a consequence, the description of the Higgs mechanism in our model of
\textit{non-perturbative} GHU makes sense only as an effective theory. It also implies that if the
bulk phase transition is of first-order then SSB persists in its vicinity.

\subsection{Monte Carlo Simulations}

Within the framework of lattice field theory, we approximate the path integral by means of Markov Chain Monte Carlo simulations. Our gauge field configurations are generated
using a Hybrid-Overrelaxation (HOR) algorithm, which consists of one heatbath step followed by $L/2$ overrelaxation steps \cite{Creutz:1987xi,Brown:1987rra},
where $L$ is the number of lattice points in the spatial direction. The heatbath algorithm is described in \cite{Kennedy:1985nu, Fabricius:1984wp} for the bulk $SU(2)$
links, and in \cite{Bunk:u1, BestFisher:vonMises} for the boundary $U(1)$ links. 

Simulating in the vicinity of a phase transition can cause critical slowing down of the updating algorithm, and can prolong the thermalisation process.
In order to minimise this effect, the first part of our thermalisation process is augmented with extra overrelaxation steps, and in extreme cases,
an additional ``magnetic field-like" term similar to \eq{eqn:P} is added to the action. It should be stressed that these techniques are employed only in the
very first thermalisation updates. They are subsequently removed, and the system is allowed to ``re-thermalise" before initiating any measurement of an observable.

For our statistical error analysis, we use the package described in \cite{Wolff:2003sm}; our quoted errors are the one-sigma statistical uncertainty
coming from this package.

\section{Phase diagram and Dimensional Reduction}\label{sec:phase_diagram}

\begin{figure}[t!]
\begin{center}
     \includegraphics[width=0.99\textwidth]{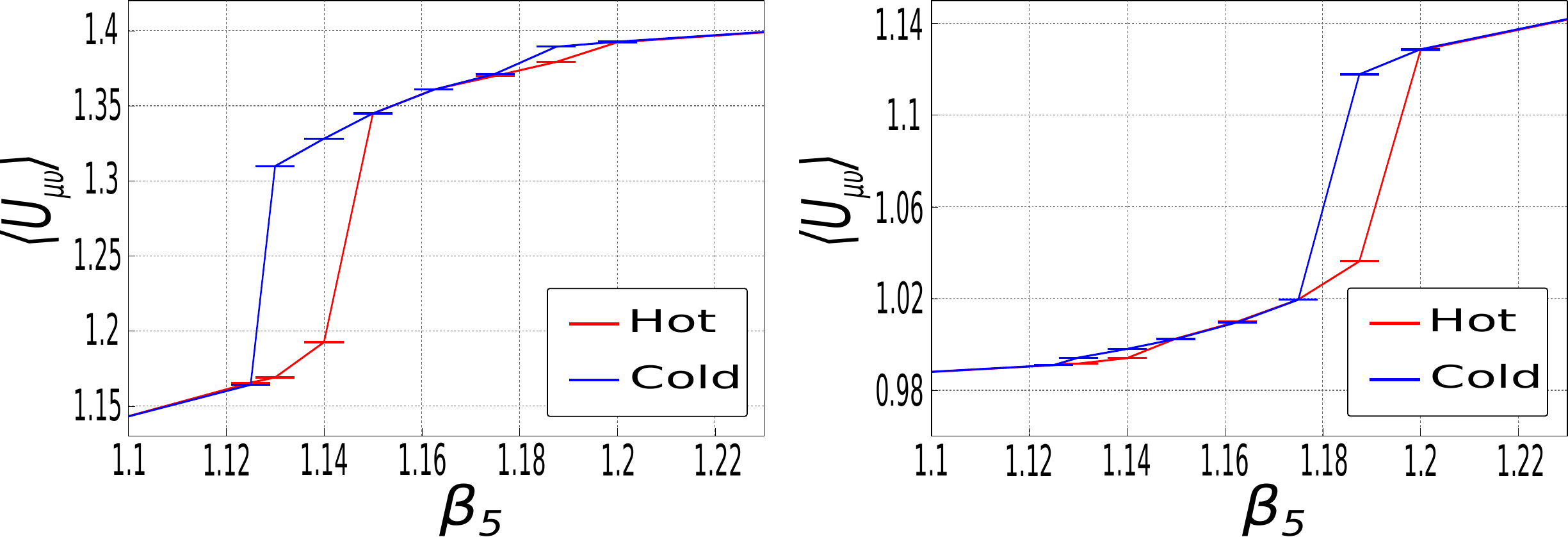}
\end{center}
    \caption{\footnotesize The left (right) panel shows example hystereses in the expectation value of the plaquette
     for fixed $\beta_4 = 1.94$ on the four-dimensional boundary (bulk) layers. The red line corresponds to a hot start and the blue line to a cold start.}
  \label{fig:hysteresis}
\end{figure}

\begin{figure}[t!]
\begin{center}
     \includegraphics[width=0.9\textwidth]{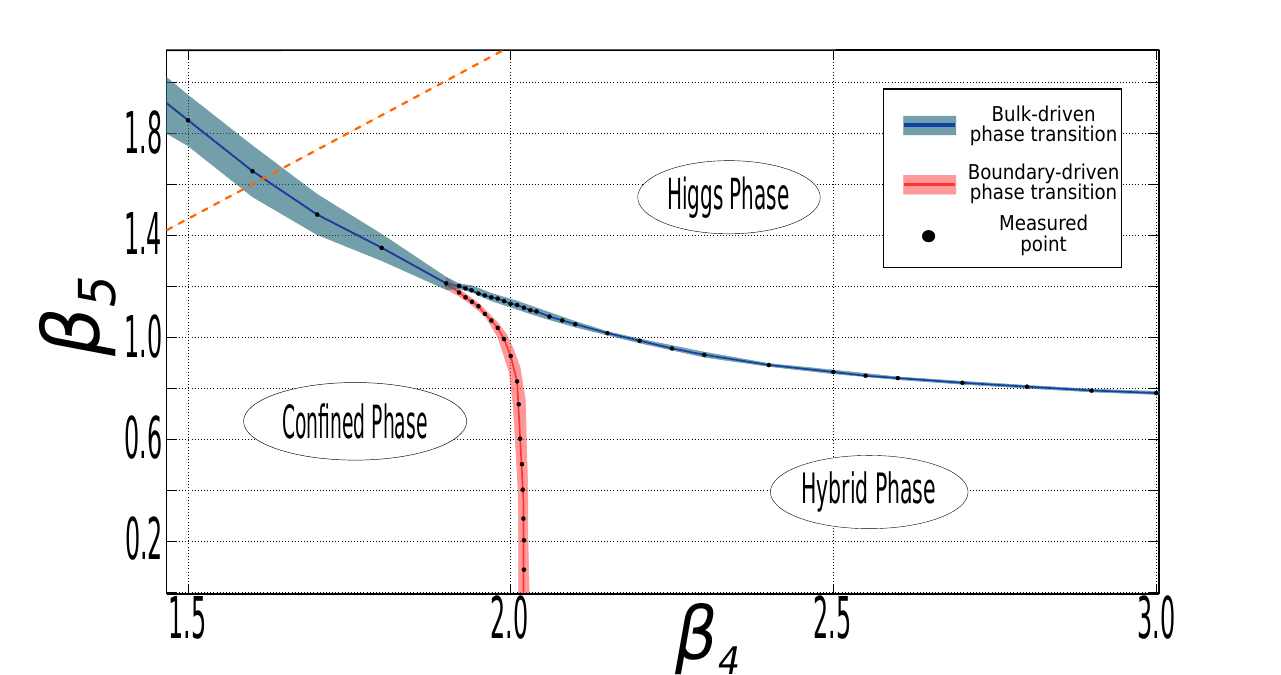}
\end{center}
    \caption{\footnotesize The phase diagram for $N_{5} = 4$ in the region of the Higgs-hybrid phase transition. The points show the location of
                         a first-order phase transition. The red and blue lines represent the width of the corresponding hystereses,
                         while the dashed orange line represents $\gamma = 1$.}
  \label{fig:phase_diag}
\end{figure}

An important step in understanding the properties of a theory is to determine the possible
phases it can exhibit. We determine the location and order of the theory's phase transitions
through the expectation value and susceptibility of plaquettes, $U_{\mu M}$, determined after
\textit{hot} and \textit{cold} starts; a hysteresis indicates a first-order phase transition,
while a smooth cross-over must be accompanied by the appropriate volume scaling of the susceptibility
in order to identify a second-order phase transition. We perform our study of the phase diagram on
lattices of various volumes $L^{4} \times 4$ and find that, once $L \geq 32$, the location and
order of the theory's phase transitions remain constant. In \fig{fig:hysteresis} we show example 
hystereses at fixed $\beta_4 = 1.94$ on the four-dimensional boundary ($n_5 = 0$) and bulk ($n_5 = 2$) layers.
In all layers we find two hystereses which correspond to the system's two phase transitions, as shown in \fig{fig:phase_diag}.  

In this study, we concentrate on the region where the bare anisotropy $\gamma \leq 1$, meaning that
the ratio of lattice spacings $a_{4}/a_{5} \leq 1$. \fig{fig:phase_diag} shows the phase diagram
for $\beta_{4} \in [1.5,3.0]$ and $\beta_{5} \in [0,2.0]$. Within this region, we find no evidence of a
second-order phase transition. Since we find only first-order phase transitions, the data points represent
the location of the centre of the corresponding hysteresis. The bands plotted through the data show
the width of the hystereses within that vicinity.

As in the case of toroidal geometry \cite{Creutz:1979dw}, we determine a first-order phase transition between the
\textit{confined} and \textit{de-confined} phases represented by the blue line in the figure. We refer to this transition as
\textit{bulk-driven} since it is the $SU(2)$ gauge field in the bulk that drives the system into a change of phase.
We follow the norm and label a phase as confined (de-confined) when the expectation value of the Polyakov loop is zero
(non-zero) in all directions. However, since massive Higgs and $Z$-like bosons have been shown to be a property of the
de-confined phase \cite{Irges:2006hg}, we refer to it as the \textit{Higgs phase}, and delay the discussion of its
properties until \sect{sec:higgs_phase}.

We determine the location of a previously unobserved first-order phase transition, represented by the red line, which
begins at $\beta_{4} = 2.02$, $\beta_{5} = 0$ and ends by merging into the bulk-driven phase transition at the \textit{triple point}
$\beta_{4} = 1.9, \beta_{5} = 1.21$. We refer to this transition as \textit{boundary-driven} since it is the $U(1)$ boundary gauge
fields that force the system into a change of phase. The four-dimensional $U(1)$ theory is known to exhibit a phase transition
at  $\beta_{U(1)} = 1.01$ \cite{Arnold:2002jk}, which corresponds directly to our value of $\beta_{4} = 2.02$ since our boundary
plaquettes take $w = 1/2$ in the orbifold action, \eq{eqn:wilson_action}, thus forcing the location of the phase transition
in terms of the four-dimensional coupling to double. This transition defines a third phase of our theory which we label as \textit{hybrid}
for reasons that will become clear when we discuss its properties in \sect{sec:hybrid_phase}.

One important observation is that the phase structure determined here agrees on a qualitative level with the one determined from
mean-field calculations in \cite{Irges:2012ih,Moir:2014aha}. Another important observation is that our phase diagram is reminiscent
of the Abelian Higgs model in four dimensions for a Higgs field of charge $q = 2$ \cite{Fradkin:1978dv,Callaway:1981rt}.
This is expected to be the theory to which the orbifold reduces in four dimensions \cite{Irges:2006hg}. In fact, the following sections
will show that we find two regions at $\gamma < 1$ where dimensional reduction occurs: the hybrid phase, and the region of the Higgs
phase close to the Higgs-hybrid phase transition. This suggests a respective mapping of the hybrid and Higgs phases of the orbifold
to the Coulomb and Higgs phases of the Abelian Higgs model. Similarly, the confined phase of the orbifold reduces on the boundary, as $\beta_{5} \rightarrow 0$,
to the confined phase of the Abelian Higgs model.  \\

\begin{table}[t!]
\begin{center}
\begin{tabular}{@{\extracolsep{0.2cm}}cccc}
\toprule
Dimension  & Yukawa & Coulomb & Confining \\
\midrule
4   &   $ c_{0} - c_{1} \frac{e^{-m_{Z} r}}{r} $  & $ c_{0} - \frac{c_{1}}{r} $        &   $ c_{0} + \sigma r - \frac{c_{1}}{r} $  \\
\\
5   &   $ c_{0} - c_{1} \frac{K_{1}(m_{Z} r)}{r} $  & $ c_{0} - \frac{c_{1}}{r^{2}} $  &   $ c_{0} + \sigma r - \frac{c_{1}}{r} $  \\
\bottomrule
\end{tabular}
\end{center}

\caption{\footnotesize Functional forms of four and five-dimensional Yukawa, Coulomb and confining potentials.
                       $m_{Z}$ corresponds to the mass of gauge boson associated with a Yukawa interaction, $\sigma$
                       corresponds to the string tension of a confining potential, $c_{i}$ correspond to constants
                       and $K_{1}$ is a Bessel function of the second kind.}
\label{tab:potentials}
\end{table}

Extra-dimensional models are only phenomenologically viable if the extra dimensions are hidden at energy scales of the Standard Model.
We are therefore only interested in regions of the phase space where dimensional reduction occurs to leave a Standard Model-like Higgs sector
on the boundary. In order to determine the dimensionality of the system for a given set of
parameters, we use the shape of the \textit{static potential}; we determine the potential, $V(r)$, between static charges of the respective gauge
group at a distance $r$ located within layers orthogonal to the fifth dimension. The potential within a given layer is extracted
from Wilson loops measured within that layer. In order to achieve an accurate determination of the potential, we replace the temporal
links of the Wilson loop by their one-link integral \cite{Parisi:1983hm} and employ the variational procedure described in \sect{sec:analysis};
in order to construct a large basis of Wilson loops we subject the spatial links to various levels of hyper-cubic
smearing (HYP) \cite{Hasenfratz:2001hp}. We adapt this smearing to the orbifold and do not consider staples extending into the temporal
or fifth dimension \cite{Yoneyamadiss}. Once we have extracted $V(r)$, we can determine its dimensionality and type by considering the global fitting forms given in \tab{tab:potentials}.
Subsequently, by considering $r^{2}F(r) \equiv r^{2}V'(r)$, which is a renormalised quantity in four-dimensions, we can examine any deviation of the potential
from the best fit in finer detail.

\subsection{The Higgs Phase}\label{sec:higgs_phase}

We now turn our attention to the Higgs phase and study its physical properties in terms of the potential it exhibits. Throughout this work,
due to their strikingly different qualitative behaviour, we will distinguish two distinct regions within the Higgs phase:
$\gamma = 1$, and $\gamma < 1$ close to the Higgs-hybrid phase transition.

From measurements of the potential along the $\gamma = 1$ line, we see no clear signs of dimensional reduction; the potential
measured within the bulk layers strongly favours a $5$-d Yukawa interpretation while on the boundary, the $4$-d Yukawa fit is no
more favoured than a $5$-d Yukawa fit. However, as we move deeper into the $\gamma < 1$ region, there is a tendency of the potential on
the boundary layers to become more $4$-d Yukawa-like, while in the bulk layers, the potential continues to strongly favour $5$-d Yukawa fits.
Significantly, once we reach low enough $\beta_{5}$, the potential on the boundary can \textit{only} be described by a $4$-d Yukawa fit. This is
strong evidence that the method of dimensional reduction is that of \textit{localisation}, that is, that the effective four-dimensional physics is localised to the boundaries.
This mechanism is reminiscent of \cite{Dvali:1996xe, Laine:2004ji}.

\begin{figure}[t!]
  \begin{center}
   \includegraphics[width=0.95\textwidth]{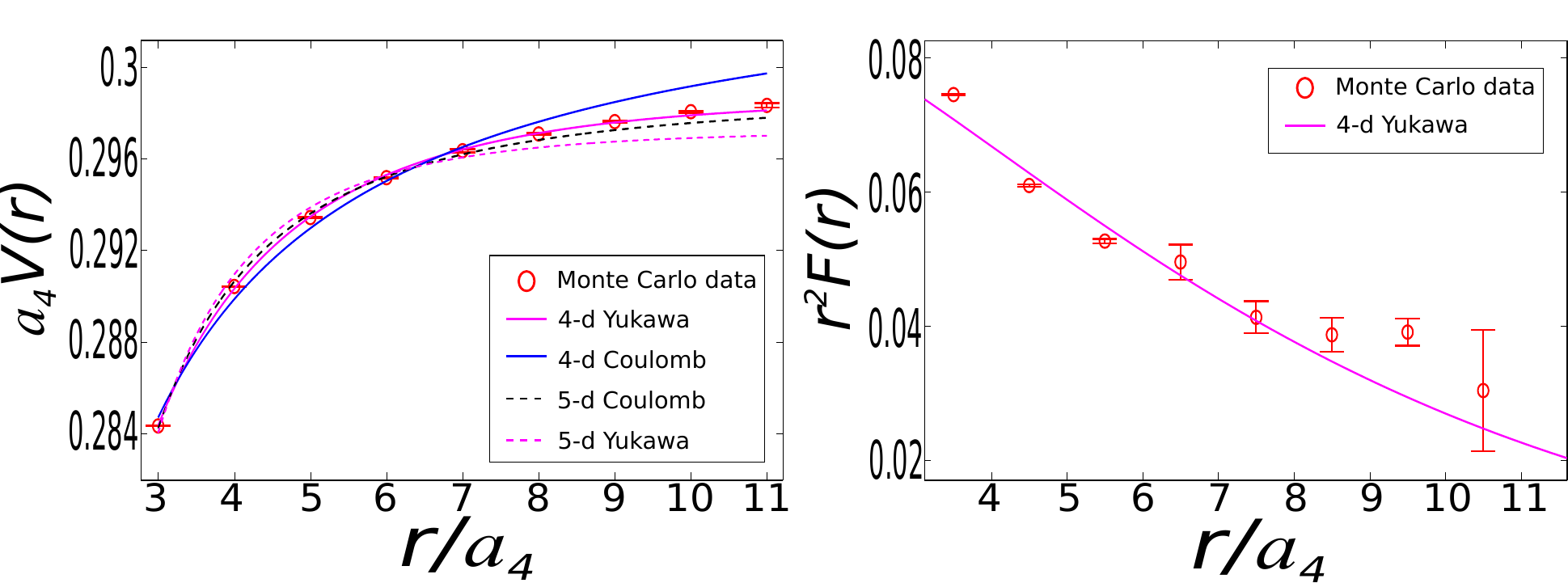}
  \end{center}
  \caption{\footnotesize The left panel shows the potential, $V(r)$, for the boundary ($n_{5} = 0$) layer calculated on a volume
                         $80 \times 32^{3} \times 4$ for $\beta_{4}=2.1$, $\beta_{5}=1.075$, which is located within the Higgs phase.
                         The right panel shows the quantity $r^{2}F(r)$ computed for the same parameter set. The measurements
                         are performed on $10,000$ gauge field configurations separated by $20$ HOR steps. The fits described in
                         the legend correspond to those in  \tab{tab:potentials}.}
  \label{fig:stat_pot}

\end{figure}

In order to illustrate the statements above, the left panel of  \fig{fig:stat_pot} shows the potential, $V(r)$, extracted from Wilson loops along the four-dimensional boundary
($n_{5} = 0$) layer for $\beta_{4}=2.1$, $\beta_{5}=1.075$. The measurements are performed on $10,000$ gauge field configurations of volume $80 \times 32^{3} \times 4$ separated by $20$ HOR steps.
The potential clearly exhibits a $4$-d Yukawa form. As shown in \tab{tab:chi2_ratios}, the \textit{goodness of fit} $\chi^{2}$ values per degree of freedom obtained for the other fits are at least an order of magnitude 
larger than that of the $4$-d Yukawa. From the fit to the $4$-d Yukawa form, we can extract the mass of the gauge boson that minimises the $\chi^{2}$ value. This turns out to be $a_{4} m_{Z} = 0.26$.
We also determine $y' = (\log(r^2F(r)))' \approx a_{4}m_{Z}$ \cite{Irges:2012ih} for the range of distances
$3 \leq r \leq 7$, and find $a_{4} m_{Z} = 0.322(66)$. Both these determinations are consistent with the value extracted from the spectroscopic calculation described in
\sect{sec:spectrum}, where we find $a_{4} m_{Z} = 0.268(3)$. It is important to note that, the $\chi^{2}$ value of the $5$-d Yukawa fit is minimised when $m_{Z} = 0$.
In this case, the $5$-d Yukawa fit trivially becomes of $5$-d Coulomb type (see \tab{tab:potentials}), and in order to distinguish the two in the figure,
we constrain the mass of the $5$-d Yukawa fit to be the one we extract from the spectroscopic calculation, namely $a_{4} m_{Z} = 0.268(3)$.

\begin{figure}[t!]
  \begin{center}
   \includegraphics[width=0.95\textwidth]{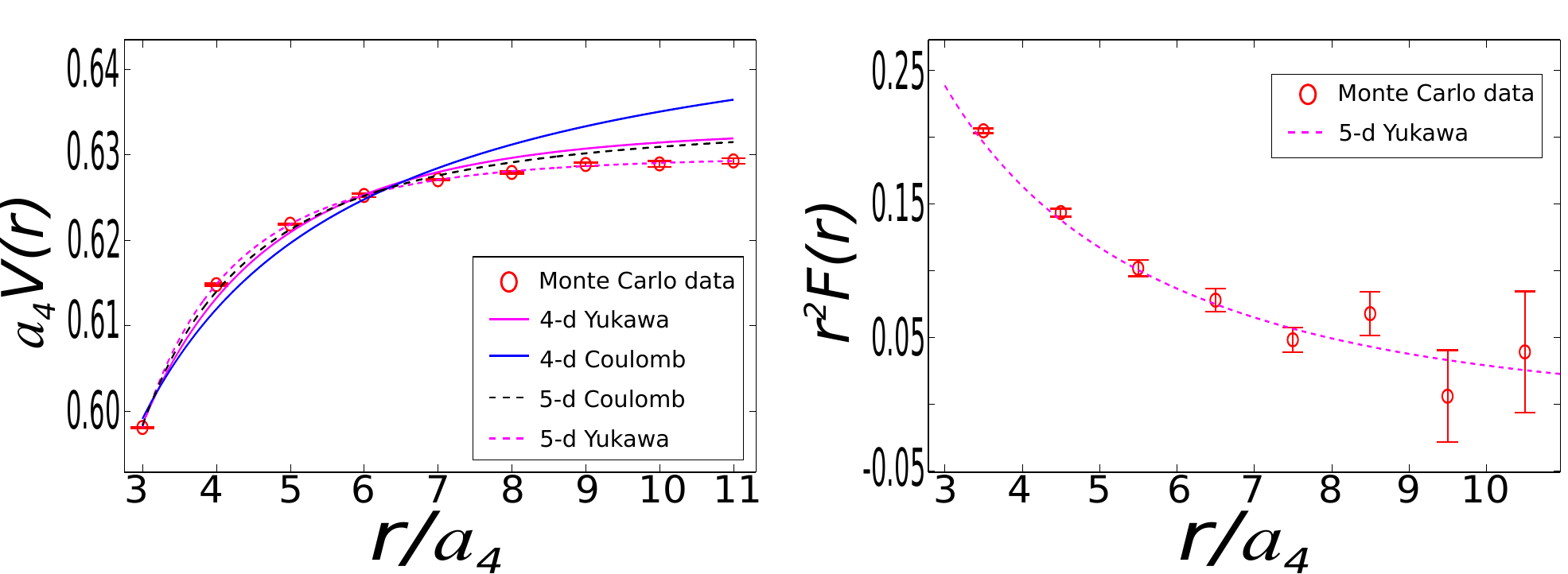}
  \end{center}
  \caption{\footnotesize The left panel shows the potential, $V(r)$, for the bulk ($n_{5} = 2$) layer calculated on a volume
                         $80 \times 32^{3} \times 4$ for $\beta_{4}=2.1$, $\beta_{5}=1.075$, which is located within the Higgs phase.
                         The right panel shows the quantity $r^{2}F(r)$ computed for the same parameter set. The measurements are
                         performed on $10,000$ gauge field configurations separated by $20$ HOR steps. The fits described in the legend
                         correspond to those in \tab{tab:potentials}.}
  \label{fig:stat_pot2}
\end{figure}

The left panel of \fig{fig:stat_pot2} shows the potential extracted from Wilson loops along the bulk ($n_{5} = 2$) layer for $\beta_{4}=2.1$, $\beta_{5}=1.075$.
The measurements are performed on the same gauge field configurations described above for the boundary potential. In this case, the potential is clearly of $5$-d Yukawa type;
the $\chi^{2}$ values of the other fits are at least two orders of magnitude larger than that of the $5$-d Yukawa. The Yukawa mass that minimises the $\chi^{2}$ value for the $5$-d Yukawa fit is
$a_{4}m_{Z} = 0.25$ which agrees with the mass we measure in our spectroscopic calculations; it seems that the SSB occurs in such a way that the mass of the gauge boson is a global property
of the system. We remark that if $m_{Z}$ is left as a free parameter in the $4$-d Yukawa fit, it yields a value which is twice as large as the spectroscopic value; in \fig{fig:stat_pot2}, we
show the $4$-d Yukawa fit with the latter mass.

The right panels of Figures \ref{fig:stat_pot} and \ref{fig:stat_pot2} show the quantity $r^{2}F(r)$ which corresponds to the potentials shown in their respective left panels. In both cases, we take the best fit
from the left panel and recast it into the form $r^{2}F(r)$. We then superimpose this fit onto the measured $r^{2}F(r)$. In both cases, we see only small deviations, if at all, of the measured $r^{2}F(r)$
from its respective fit. Therefore, the correction to the boundary potential coming from the five-dimensional bulk is negligible, confirming that dimensional reduction
has taken place on the boundary via localisation.

\subsection{The Hybrid Phase}\label{sec:hybrid_phase}

By sufficiently decreasing $\beta_{5}$ and keeping $\beta_{4} > 2.02$, we enter a third phase of the theory. We label it as hybrid
due to the fact that we observe four-dimensional de-confined $U(1)$ theories on the boundaries, and four-dimensional
confined $SU(2)$ theories within the bulk layers; the layers in this phase appear to behave as independent four-dimensional slices that
are very weakly coupled and hence, the physical content of a given layer is completely dominated by the gauge group of that layer. Furthermore,
the Polyakov loop along the fifth dimension has zero expectation value, whereas the expectation value of Polyakov loops winding in the other four
dimensions is zero in the bulk and non-zero on the boundary. This behaviour is strongly reminiscent of the \textit{layered phase} described in \cite{Fu:1983ei}.
\begin{figure}[t!]
  \begin{center}
    \includegraphics[width=0.95\textwidth]{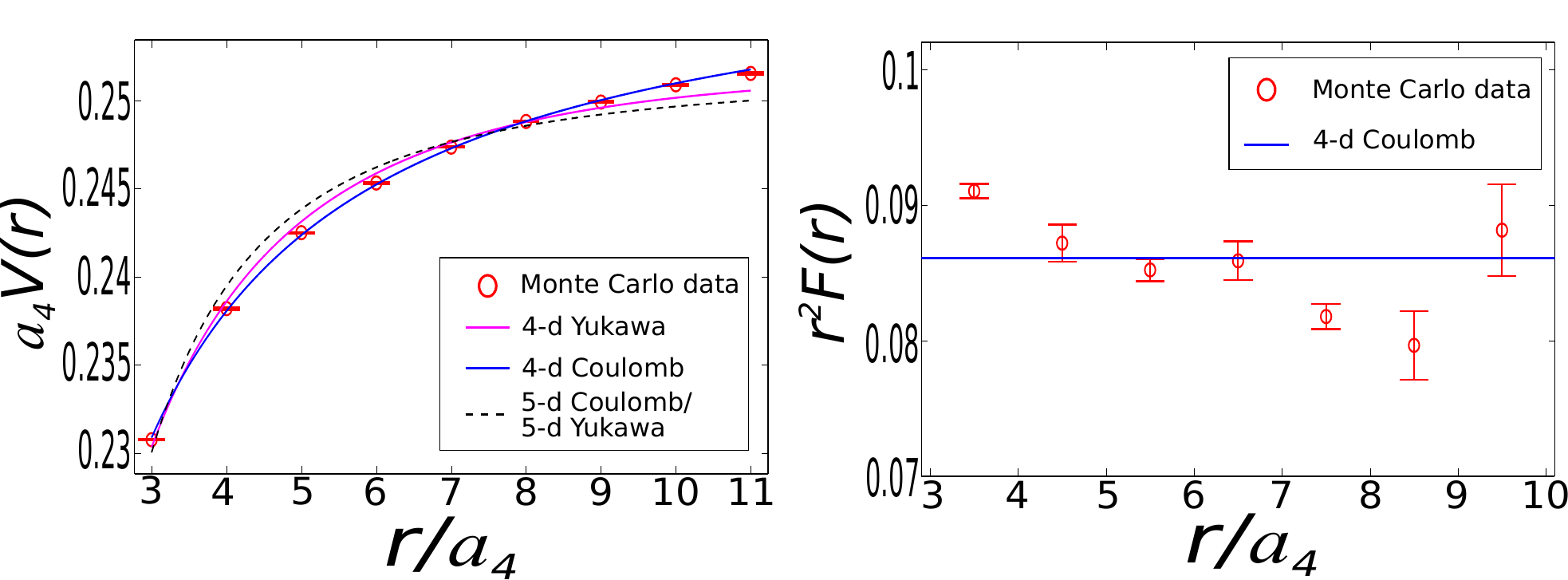}
  \end{center}
  \caption{\footnotesize The left panel shows the potential, $V(r)$, on the boundary ($n_{5} = 0$) layer calculated on a volume
                         $80 \times 32^{3} \times 4$ for $\beta_{4}=2.6$, $\beta_{5}=0.6$, which is located within the hybrid phase.
                         The right panel shows the quantity $r^{2}F(r)$ computed for the same parameter set. The measurements are
                         performed on $5,000$ gauge field configurations separated by $5$ HOR steps. The fits described in the legend
                         correspond to those in \tab{tab:potentials}.}
  \label{fig:stat_pot_hybrid}
\end{figure}

\begin{figure}[t!]
  \begin{center}
    \includegraphics[width=0.95\textwidth]{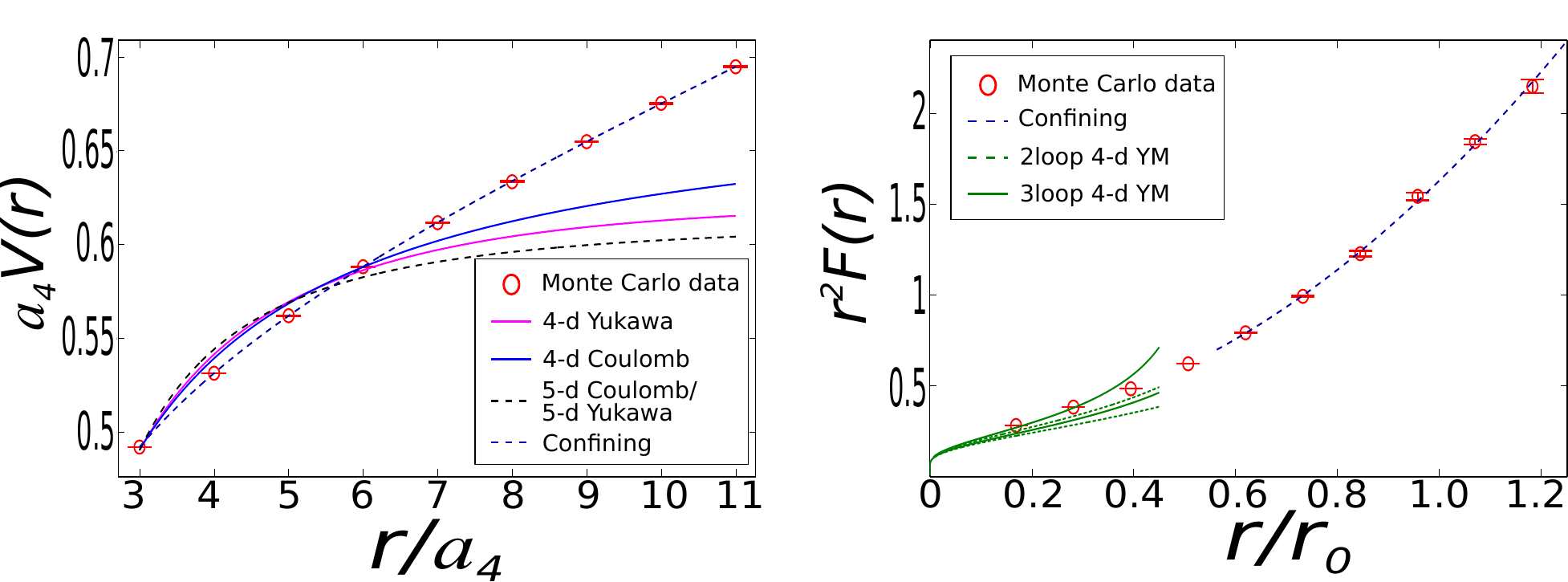}
  \end{center}
  \caption{\footnotesize  The left panel shows the potential, $V(r)$, within the bulk ($n_{5} = 2$) layer calculated on a volume
                          $80 \times 32^{3} \times 4$ for $\beta_{4}=2.6$, $\beta_{5}=0.6$, which is located within the hybrid phase.
                          The right panel shows the quantity $r^{2}F(r)$ computed for the same parameter set, in units of the
                          Sommer scale $r_{0}$. The measurements are performed on $5,000$ gauge field configurations separated by $5$ HOR steps.
                          The green lines (dashing) show the three-loop (two-loop) calculations of $r^2 F(r)$ in four-dimensional pure gauge
                          $SU(2)$ theory. The fits described in the legend correspond to those in \tab{tab:potentials}.}
  \label{fig:force_SU2_hybrid}
\end{figure}

In order to validate these claims, we show in the left panel of \fig{fig:stat_pot_hybrid}, the potential, $V(r)$, extracted from Wilson loops measured along the boundary ($n_{5} = 0$) layer for $\beta_4=2.6$, $\beta_5=0.6$. From the figure it is clear that the potential can only be described by a $4$-d Coulomb fit. As shown in \tab{tab:chi2_ratios}, the $\chi^{2}$ values for the other fits are two orders of magnitude larger than that of the $4$-d Coulomb. On the other-hand, the left panel of \fig{fig:force_SU2_hybrid} shows the potential extracted from Wilson loops along the $n_{5} = 2$ bulk layer. In this case we find that only a confining string-like potential can describe the data; the other fits have $\chi^{2}$ values many orders of magnitude larger than that of the confining. Since we obtain confining potentials in the bulk layers, we can assume that the $-c_{1}/r$ term of the confining fit is the L\"{u}scher term, $-(d-2)\pi/(24r)$ \cite {Luscher:1980fr,Luscher:1980ac}, giving us a probe of the dimensionality, $d$, of a given layer. For the potential shown in the left panel of \fig{fig:force_SU2_hybrid}, we obtain a value of $d = 3.97$. Furthermore, we compute the slope, $c_{1}(r) = r^{3}F'(r)/2$ \cite{Luscher:2002qv}, which we find to plateaux at a value of $-0.288(16)$. This is consistent with the expected value
of $-\pi/12$ for a $4$-d confining theory, emphasising the fact that the potential within a given layer of the bulk is indeed four-dimensional.

When we attempt to compute the spectrum of the scalar channel within the hybrid phase, using the Higgs-type operators described in \sect{sec:operators}, we can not determine a particle mass. Instead, the signal coming from correlation functions almost immediately drop below
machine precision signalling that the mass decouples from the theory at the scale of the hybrid phase; the bulk behaves like a two-colour
version of QCD and we expect that the scale here is the Lambda parameter of $SU(2)$. This expectation is justified since
we can compute the Sommer scale $r_{0}$ \cite{Sommer:1993ce}, which is defined as a typical physical scale ($0.5$ fm) of QCD, from the potential shown in the left panel of \fig{fig:force_SU2_hybrid}. We find that we reach the Sommer scale when the distance between two static charges is $8.87(5)$
in lattice units, meaning that the bulk layers within the hybrid phase are sensitive to physics at the scale of two-colour QCD. Since we can not determine a scalar mass using our interpolating operators, and we know that glueball-like particles should be observable at the scale of two-colour QCD, we conclude that our operators do not have any significant overlap with glueball-like states and thus only probe the degrees of freedom relevant for Higgs-like particles, in agreement with \cite{Philipsen:1997rq}. We remark that the behaviour of the scalar and gauge boson masses, in the vicinity of the Higgs-hybrid phase transition, is consistent with the spectrum computed within the four-dimensional Abelian Higgs model \cite{Evertz:1986ur}.

\begin{table}[t!]
\small
\begin{center}
\begin{tabular}{lccccc}
\toprule
Phase (Layer)  & 4-d Yukawa & 5-d Yukawa & 4-d Coulomb & 5-d Coulomb & Confining \\
\midrule
Higgs (Boundary)   &  $11.9$ & $112.7$ & $1.0\times10^3$ & $138.0$ & -  \\
Higgs (Bulk)       &  $918.7$ & $2.1$ & $4.0\times10^3$ & $361.8$ & -  \\
\midrule
Hybrid (Boundary)   &  $208.9$ & $698.9$ & $9.9$ & $692.1$ & -  \\
Hybrid (Bulk)       &  $1.6\times10^5$ & $2.4\times10^5$ & $9.7\times10^4$ & $2.4\times10^5$ & 2.2  \\
\bottomrule
\end{tabular}
\end{center}
\caption{\footnotesize $\chi^2$ values per degree of freedom for the fits to the static potentials shown in Figures \ref{fig:stat_pot}, \ref{fig:stat_pot2}, \ref{fig:stat_pot_hybrid} and \ref{fig:force_SU2_hybrid}.}
\label{tab:chi2_ratios}
\end{table}

By computing the quantity $r^2 F(r)$, we can get a handle on the correction to the physics along the four-dimensional bulk layers coming from the extra dimension. The right panel of \fig{fig:force_SU2_hybrid} shows the quantity $r^2 F(r)$ at distances in units of the Sommer scale $r_{0}$. One reaches perturbative scales as $r/r_{0} \rightarrow 0$, and the green lines (dashing) correspond to three-loop (two-loop) calculations of $r^2 F(r)$ in four-dimensional pure gauge $SU(2)$ theory, see \cite{Knechtli:2011gq}. The fact that we see no deviation of our calculation from perturbative predictions down to very small values of $r/r_{0}$ tells us that the correction to the four-dimensional physics is negligible and occurs at energy scales much larger than the layers in
the bulk are sensitive to. Furthermore, by recasting the confining fit shown in the left panel of \fig{fig:force_SU2_hybrid} into the form $r^2 F(r)$ and superimposing it onto our calculated $r^2 F(r)$, we see that there is no significant deviation from the $4$-d confining fit. Therefore, it is clear that the correction to the bulk layers coming from the extra dimension is negligible. In a similar fashion, the right side of
\fig{fig:stat_pot_hybrid} shows that the correction to the $4$-d Coulomb boundary layers coming from the extra dimension is negligible.

Significantly, on any four-dimensional layer within the hybrid phase, the $4$-d Yukawa fit yields a zero mass for the gauge boson when the parameter $m_{Z}$ is free. However, when $m_{Z} = 0$, the $4$-Yukawa fit trivially becomes of $4$-d Coulomb type. Therefore, in order to distinguish the two in \fig{fig:stat_pot_hybrid}, we fix the mass in the $4$-d Yukawa fit to be $a_{4} m_{Z} = 0.2$, as it is a typical value we obtain
for the $Z$ boson mass in the lower region of the Higgs phase. The fact that $m_{Z} = 0$ minimises the $\chi^{2}$ of the $4$-d Yukawa fit
is strong evidence that there is no massive gauge boson and hence, no SSB in the hybrid phase.

\section{Spectroscopic Extraction}\label{sec:spec_extract}
We obtain spectral information from two-point correlation functions
\begin{equation}\label{eqn:correlation_function}
C_{ij}(t) = \langle \mathcal{O}_{i}(t) \mathcal{O}^{*}_{j}(0) \rangle  - \langle \mathcal{O}_{i}(t) \rangle \langle \mathcal{O}^{*}_{j}(0) \rangle   ~,
\end{equation}
where $\mathcal{O}^{*}_{j}(0)$ and $\mathcal{O}_{i}(t)$ are interpolating fields which correspond in the Hamiltonian formalism to operators that create
and annihilate a state respectively. After inserting a complete set of eigenstates of the Hamiltonian, the two-point function becomes
\begin{equation}
C_{ij}(t) = \sum_{n} \frac{Z^{n *}_{i} Z^{n}_{j}}{2E_{n}} e^{-E_{n}t}~,
\end{equation}
where $E_{n}$ is the energy of the $n^{th}$ state and the sum is over a discrete set of states due to the finite volume.
The vacuum-state matrix elements $Z^{n *}_{i} \equiv \langle n | \mathcal{O}^{\dagger}_{i} | 0 \rangle$ can be interpreted as a
measure of the contribution of an operator to a given state (see \sect{sec:analysis}).

\subsection{Higgs and $Z$ Boson Operators \label{sec:operators}}

The main aim of this study is to reliably determine the low-lying spectrum of the
$J^{PC} = 0^{++}$ (Higgs) and $1^{--}$ ($Z$) channels within the Higgs phase.
The quality of our spectral determination relies heavily on our ability to
probe the relevant degrees of freedom of the theory. In order to achieve this in the Higgs channel,
we follow \cite{Irges:2006hg}, and begin by defining an extra-dimensional gauge potential.
If the extra dimension is a discrete circle, it is given up to O($a^{3}$) by
\begin{equation}
v(n_{\mu}) = \frac{1}{4N_{5}} (p(n_{\mu}) - p^{\dagger}(n_{\mu}))~,
\end{equation}
where $p(n_{\mu})$ is the Polyakov loop winding around the extra dimension. In order to construct
this potential on the orbifold, $S^{1}/\mathbb{Z}_{2}$, the orbifolding condition \eqref{eqn:orbifold_condition}
is imposed on the Polyakov loop, resulting in
\begin{equation}\label{eq:p}
p(n_{\mu}) = l(n_{\mu}) g l^{\dagger}(n_{\mu}) g^{\dagger}~,
\end{equation}
where $l(n_{\mu})$ is the Polyakov line extending across the extent of the extra dimension. Taking the potential,
$v(n_{\mu})$, as our basic building block, we define the scalar field, $h(n_{\mu})$, as
\begin{equation}\label{eqn:h}
h(n_{\mu}) = [ v(n_{\mu}), g ]~.
\end{equation}
Since $h(n_{\mu})$ is traceless, we define 
\begin{equation}\label{eqn:H}
\mathcal{H}(n_{0}) = \frac{1}{L^{3}}\sum_{n_{k}} \tr[ h(n_{\mu}) h^{\dagger}(n_{\mu}) ]~,
\end{equation}
where the sum is over all spatial directions $k$, as our primary Higgs operator. From its construction, it is clear that
it will act as a direct probe of the theory's Higgs-like degrees of freedom. 

On the lattice, it is well-known that the use of the variational approach described in \sect{sec:analysis}
leads to significantly improved spectral determinations. However, it requires the use of a reasonably large basis
of interpolating operators, and hence, we add to our basis in the Higgs channel, a second type of operator
\begin{equation}\label{eqn:P}
\mathcal{P}(n_{0}) = \frac{1}{L^{3}} \sum_{n_{k}} \tr[ p(n_{\mu}) ]~ .
\end{equation}
We further increase our basis through the combined use of two smearing procedures. Firstly, we smear the gauge links $i$ times
using HYP smearing. We then construct Polyakov loops, $p_{i}( n_{\mu} )$, from the HYP smeared links $U_{i}$,
and apply $j$ levels of an APE-like smearing to them
\begin{equation}\label{eqn:p_ij}
p_{ij}(n_{\mu}) = (1 - c) p_{i(j-1)}(n_{\mu}) + \frac{c}{6} \sum_{\substack{n_{0}=n'_{0}\\ |\vec{n} - \vec{n}'| = a_{4}}} U_{i}(n_{\mu}; n'_{\mu}) p_{i(j-1)}(n'_{\mu}) U^{\dagger}_{i}(n_{\mu}; n'_{\mu})~,
\end{equation}
where we take the constant $c = 0.7$, and the link $U_{i}(n_{\mu}; n'_{\mu})$ connects co-ordinate $n_{\mu}$ to $n'_{\mu}$. Using these smeared Polyakov loops as the basic ingredients of equations \eqref{eqn:H} and \eqref{eqn:P},
we arrive at our final operator basis for the Higgs channel, namely $\mathcal{P}_{ij}$ and $\mathcal{H}_{ij}$. \\

\begin{table}
\small
\begin{center}
\begin{tabular}{@{\extracolsep{0.2cm}}|c c c|}
\hline
$\mathcal{O}$ & $J^{PC}$ & ${\mathcal S}$ \\
\hline
$\mathcal{H}$              & $0^{++}$ & $+$ \\
$\mathcal{P}$              & $0^{++}$ & $+$ \\
\hline
$\mathcal{Z}$          & $1^{--}$ & $-$ \\
$\mathcal{Z}'$         & $1^{--}$ & $-$
\\
\hline
\end{tabular}
\end{center}
\caption{\label{tab:stick}\footnotesize The quantum numbers of our Higgs and $Z$ boson interpolating operators.
$J^{PC}$ denotes the total angular momentum, parity and charge conjugation in the continuum, while ${\mathcal S}$ refers
to their parity under the stick symmetry. }
\end{table}

Due to the nature of the theory's SSB, we expect to determine a massive $Z$-like vector boson.
Since it is the order parameter for SSB and possesses the correct quantum numbers, the obvious choice of probe
for this particle is the vector Polyakov loop. We define it as
\begin{equation}\label{eqn:Z}
\mathcal{Z}(n_{0}, k)  =  \frac{1}{L^{3}} \sum_{n_{1}, n_{2}, n_{3}}  \tr [ g U (n_{\mu}, k) \alpha(n_{\mu} + a_{4}\hat{k}) U^{\dagger}(n_{\mu}, k) \alpha(n_{\mu}) ]~,
\end{equation}
where $\alpha(n_{\mu})$ is the $SU(2)$ projection of \eq{eqn:h}. As
in the case of the Higgs channel, we can significantly improve our spectral determinations through the use of a large
basis of interpolating operators. We therefore define a second type of operator, which we denote as $\mathcal{Z'}$ due to its
importance in determining an excited $Z'$-like state, by
\begin{equation}\label{eqn:Z'}
\mathcal{Z'}(n_{0}, k) = \frac{1}{L^{3}} \sum_{n_{1}, n_{2}, n_{3}} \tr [ g  U(n, k) l(n_{\mu} + a_4\hat{k})  U^{\dagger}(n', k) l^{\dagger}(n_{\mu})]~,
\end{equation}
where $n=(n_{\mu},n_{5}=0)$ and $n'=(n_{\mu},n_{5}=N_5)$. Using the smeared Polyakov loops
of \eq{eqn:p_ij}, we arrive at our final operator basis for the $Z$ channel, namely $(\mathcal{Z}_{k})_{ij}$ and $(\mathcal{Z'}_{k})_{i}$.

In \tab{tab:stick}, we show the continuum quantum numbers of our interpolating operators along with their parities under the stick symmetry $\mathcal{S}$. 
Note that, in practice we construct our Higgs and $Z$ boson operators at $n_{5}=0$, since we find that in general,
$\langle \mathcal{O} \rangle = \langle \mathcal{F}[\mathcal{O}] \rangle$ holds.

\subsection{Analysis of Two-Point Correlation Functions \label{sec:analysis}}

In order to robustly determine spectral quantities, we employ a well-established variational
technique \cite{Michael:1985ne,Luscher:1990ck}. Within a given symmetry channel, $\Lambda$, this amounts to
constructing an $N_{\Lambda} \times N_{\Lambda}$ matrix of two-point correlation functions, and solving the
generalised eigenvalue equation   
\begin{equation}
C_{ij}(t)v^{n}_{j} = \lambda^{n}(t,t_{0})C_{ij}(t_{0})v^{n}_{j}~,
\end{equation}
where $C_{ij}(t)$ is given by \eq{eqn:correlation_function}, and $n = 1, 2, \dots , N_{\Lambda}$.
Here, $N_{\Lambda}$ refers to the number of interpolating operators used in a given channel. The solution procedure
yields a set of eigenvalues $\{ \lambda^{n}(t,t_{0}) \}$, ordered such that
$\lambda^{1}(t,t_{0}) > \lambda^{2}(t,t_{0}) > \dots > \lambda^{N_{\Lambda}}(t,t_{0})$, and a corresponding set of
eigenvectors $\{v^{n}_{j}\}$ for each $t$ and reference time-slice $t_{0}$. At large enough values of $t$,
the eigenvalues are proportional to $e^{-E_{n}(t-t_{0})}$ and hence, the energy, $E_{n}$, of state $n$ can be
extracted via
\begin{equation}
a_{4}m_{n} \equiv a_{4}E_{n} = \ln \left( \frac{\lambda^{n}(t, t_{0})}{\lambda^{n}(t + a_{4}, t_{0})} \right)~,
\end{equation}
which is known as the \textit{effective mass} of state $n$.

\begin{figure}[t]
  \begin{center}
  \includegraphics[width=0.85\textwidth]{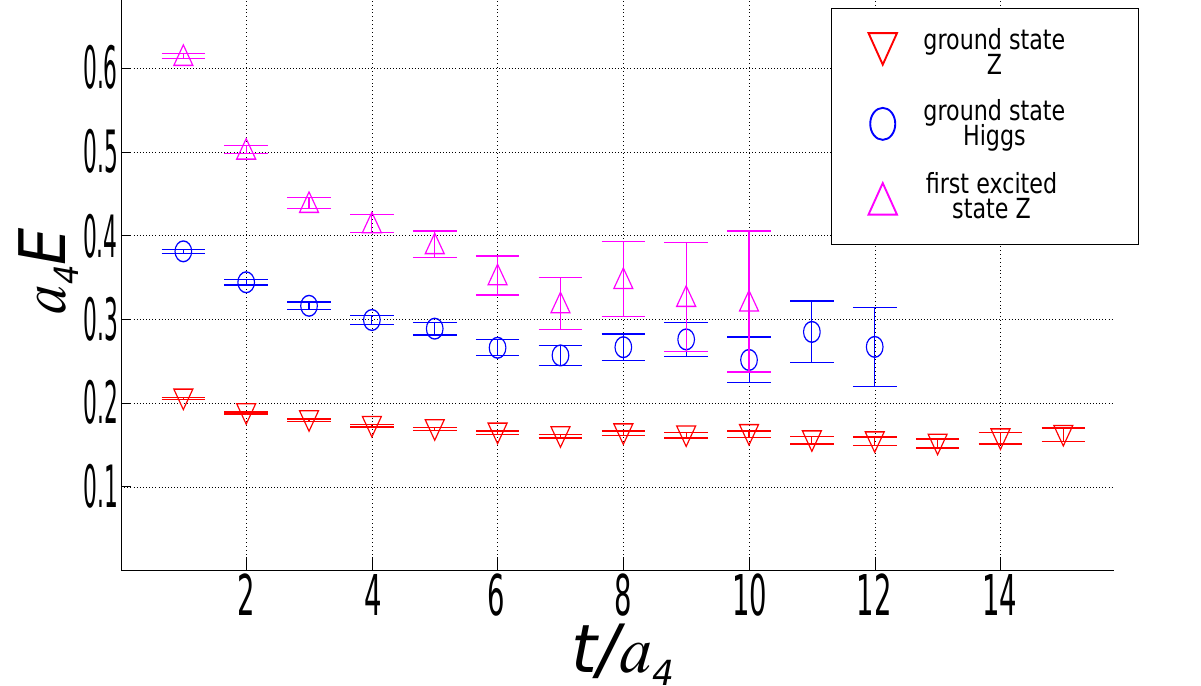}
  \end{center}
  \caption{\footnotesize{Effective masses of the ground states in the Higgs and $Z$ boson channels along with the first excited state in the $Z$ channel. 
                         The measurments are obtained using $5,000$ gauge field configurations, each separated by $5$ HOR steps. 
                         The parameter set for this calculation is $\beta_{4} = 3.0, \beta_{5} = 0.83$, $N_{5} = 4$, $N_{Higgs} = 12$ and $N_{Z} = 8$}.}
  \label{fig:eff_mass}
\end{figure}

In order to ensure that we determine reliable effective mass values, we take the following steps. Firstly, we construct an \textit{optimised}
operator basis in each symmetry channel. We achieve this by combining the \textit{operator overlap} strategy described
in \cite{Dudek:2010wm}, which uses vacuum-state matrix elements to distinguish between operators that significantly contribute to a given
state from those that do not, with the \textit{absolute overlap} strategy defined in \cite{Knechtli:2000df}, which gives an overall measure
of how well an operator basis can reproduce a given eigenstate of the Hamiltonian. In general, we find that the addition of subsequent operators to
our optimised bases do not significantly alter effective mass values of the low-lying spectrum.

Secondly, we only determine effective mass values where a \textit{distinct plateaux} between time-slices $t_{i}$ and $t_{f}$ is evident.
By fitting the function $c + e^{-\delta_{m} t}$ to the effective mass, we define $t_{i}$ to be the time-slice for which the exponential term
of the fit is less than the one-sigma statistical uncertainty of the effective mass.
We define $t_{f}$ to be the time-slice for which the value of the effective mass at $t_{f}$ is no longer within one-sigma of the fit function, or the one-sigma statistical uncertainty of the effective mass is twice that of $t_{i}$. Furthermore, we only accept a plateaux that extends for at least four time-slices. 

In \fig{fig:eff_mass}, we show the results of applying the above procedure in both the Higgs
and $Z$ channels for the parameter set $\beta_{4} = 3.0, \beta_{5} = 0.83$, $N_{5} = 4$, and optimal bases $N_{Higgs} = 12$
and $N_{Z} = 8$. The effective masses shown are obtained using $5,000$ gauge field configurations separated by $5$ HOR steps, and are typical of all our calculations.
It is clear that we reliably determine the ground state in both the Higgs and $Z$ channels, and the first excited state in the $Z$ channel.
Although we see some evidence that an excited state in the Higgs channel may be present within the range $0.7 \leq a_{4}E \leq 0.9$,
the effective mass does not meet the criteria described above and we do not show it in \fig{fig:eff_mass}.

\section{Spectrum}\label{sec:spectrum}

We now present the results of our spectroscopic calculations; preliminary results have already been presented in \cite{Moir:2014aha}.
Since we are interested in the properties of the theory where SSB occurs, we perform our spectroscopic calculations within the Higgs phase.
Within the Standard Model, the ratio of the Higgs mass \cite{ATLAS:2012gk,CMS:2012gu} to that of the $Z$, $\rho \equiv m_{H} / m_{Z} \approx 1.38$.
In order for  Gauge-Higgs Unification scenarios to be phenomenologically viable, they should achieve a similar value for a range of
physically similar phase space parameters; an indication that this is achievable in our current $SU(2)$ model would encourage further
explorations into larger models that can account for all the degrees of freedom within the Higgs sector.

In this section we fix $N_{5} = 4$, and in \sect{sec:changing_n5} we take first steps in exploring the effect of varying the
size of the extra dimension. Where instructive, we show our results in units of the energy scale $1/R$, where $R$ is the radius of
the extra dimension. However, it must be noted that, once an anisotropy is introduced into the couplings, the radius is only determined classically.
In order to overcome this limitation, the renormalised anisotropy must be obtained, and work in this direction is already under way.

Here we explore two distinct regions. The first being that of isotropic couplings, where $\beta \equiv \beta_{4} = \beta_{5}$,
and the second being that of anisotropic couplings such that $\gamma < 1$. We show all of our spectroscopic results for these regions in Appendix \ref{app:eff_masses}.

\subsubsection*{\center{\bm{$\underline{\gamma = 1}$}}}\label{sec:spectrum_isotropic}

\begin{figure}[t!]
  \begin{center}
   \vspace*{-1.5cm}\includegraphics[width=0.95\textwidth]{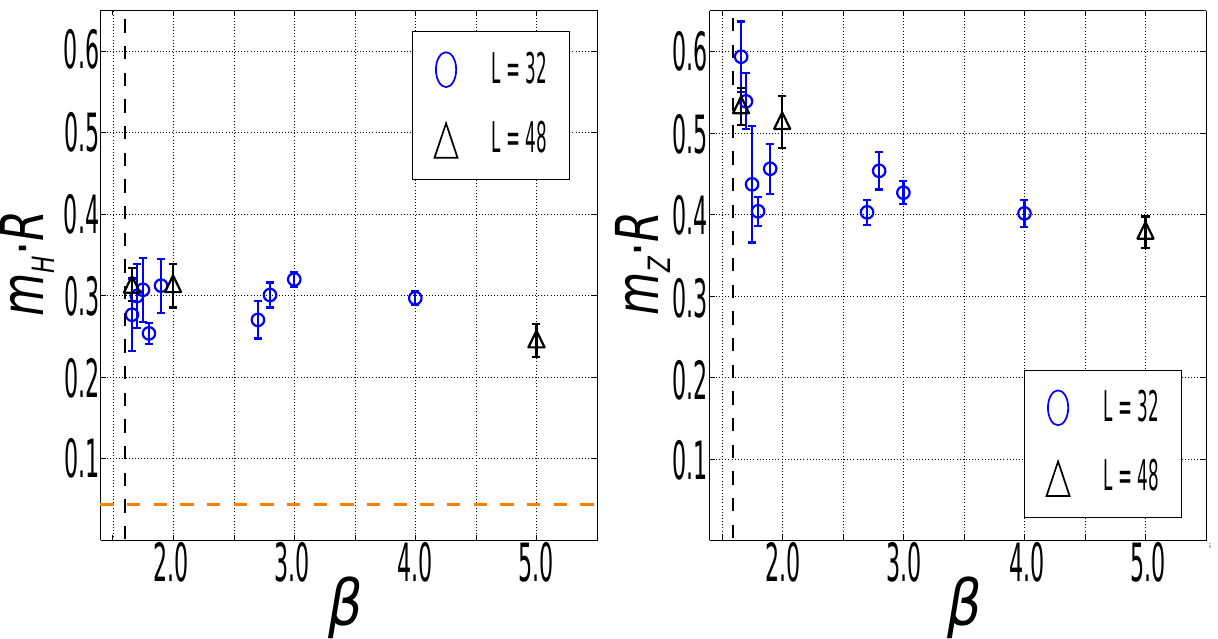}
  \end{center}
  \caption{\footnotesize The Higgs (left panel) and $Z$ (right panel) boson 
masses, in units of $1/R$,
determined for various values of the isotropic coupling, $\beta$, and 
volumes $80 \times L^{3} \times 4$.
The black vertical line corresponds to the first-order phase transition 
at $\beta \approx 1.65$ between the
confined and Higgs phases. The orange horizontal line refers to the value 
of the one-loop Higgs mass from perturbation theory 
\cite{vonGersdorff:2002as,Cheng:2002iz}.}
  \label{fig:spectrum_iso}
\end{figure}

In \fig{fig:spectrum_iso}, we show our calculated Higgs and 
$Z$ boson masses, in units of $1/R$,
for various values of the isotropic coupling, $\beta$, and volumes, 
$80 \times L^{3} \times 4$; within all of our spectroscopic
calculations, we fix the temporal extent $T = 80$. It is clear that we 
see no significant volume dependence for $L \geq 32$
and consequently, unless otherwise stated, we perform our calculations 
on volumes $80 \times 32^{3} \times N_{5}$. Significantly, we find $a_{4}m_{Z} \neq 0$ as $L \rightarrow \infty$ and hence, confirm the existence of SSB within the Higgs phase.
The black vertical line represents the bulk-driven Higgs-confined phase 
transition. The
orange horizontal line represents the Higgs mass calculated from a 
one-loop calculation in perturbation theory 
\cite{vonGersdorff:2002as,Cheng:2002iz},
making it clear that the mechanism giving rise to the masses is highly non-perturbative and different from a finite
temperature Debye mass, as discussed in \sect{sec:ssb}.  Clearly, for our range of $\beta$ values, $m_{H} < m_{Z}$. However, in 
the perturbative limit, where $\beta \rightarrow \infty$, we expect the
 Higgs mass to approach its one-loop value and the $Z$ to become massless, 
implying that a flip in the hierarchy is guaranteed. 
For our explored values of $\beta$, we find $\rho$ to be within the range 
$0.55 \leq \rho \leq 0.75$,
suggesting that one would have to go to extremely large (and computationally impractical) 
values of $\beta$ to achieve a value of $\rho \approx 1.38$,
if at all possible \footnote{As we discussed at the end
of \sect{sec:ssb}, the continuum limit at $\beta\to\infty$ is trivial.
At finite values of $\beta$ we only find first order phase transitions.
Therefore the theory has to be treated as an effective theory at finite 
values of the cutoff.}.
We therefore rule out the possibility, at least for 
$N_{5} = 4$, of achieving a Standard
Model-like hierarchy when $\gamma = 1$.

\subsubsection*{\center{\bm{$\underline{\gamma < 1}$}}}\label{sec:spectrum_low_gamma}

We now turn our attention to the case of anisotropic couplings resulting in 
$\gamma < 1$. The mean-field calculation
presented in \cite{Irges:2012mp} suggest that our theory should produce a 
Standard Model-like hierarchy close to
the bulk-driven phase transition. Furthermore, it suggests that an excited 
$Z'$ particle is present in the spectrum and that it
is possible to construct renormalized trajectories along the phase diagram 
such that the physical quantities remain
constant as the lattice spacing is varied.

\begin{figure}[t]
  \begin{flushleft}
    \includegraphics[width=0.95\textwidth]{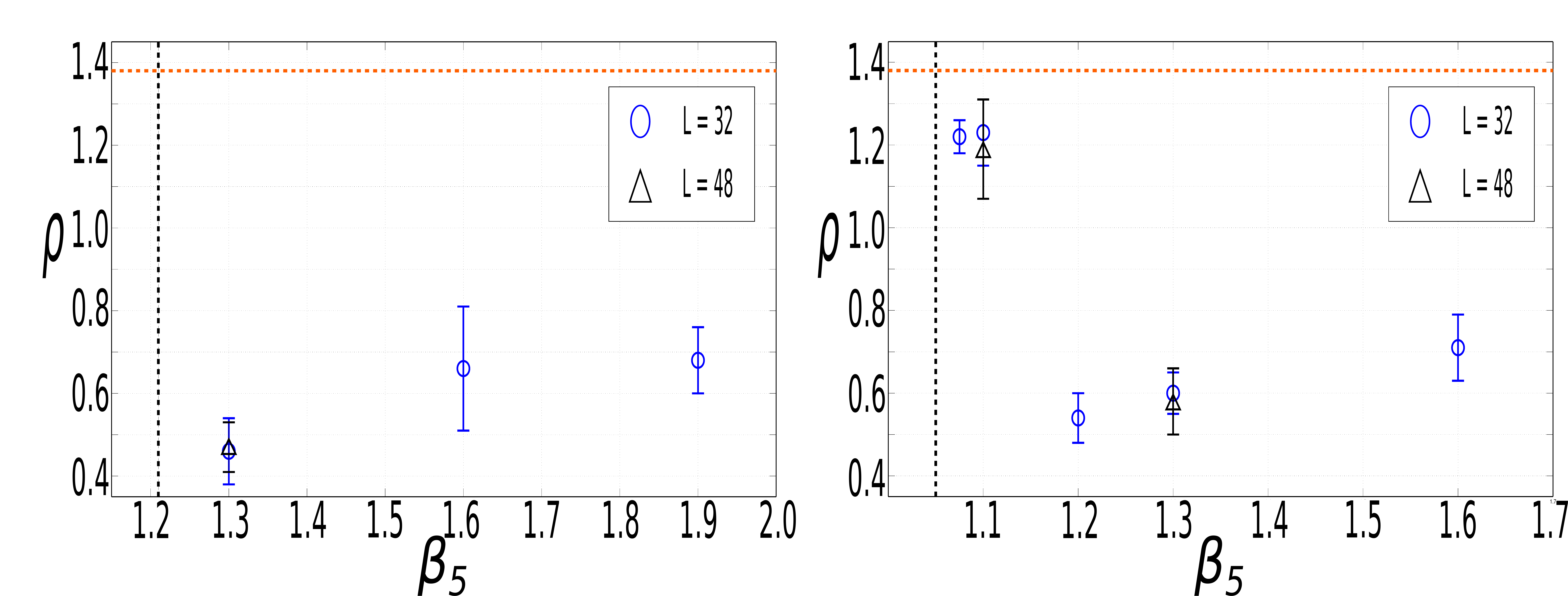}
  \end{flushleft}
  \caption{\footnotesize The left (right) panel shows the ratio $\rho = m_{H}/ m_{Z}$ for various values of $\beta_{5}$
and fixed $\beta_{4} = 1.9$ ($2.1$). The calculations are performed on various volumes, $80 \times L^{3} \times 4$, as described by the legend. 
The black vertical line on the left (right) panel shows the location of the Higgs-confined (Higgs-hybrid) phase transition,
while the orange horizontal line shows the Standard Model value for $\rho$.}
  \label{fig:b4_19_21}
\end{figure}

In order to explore this region, we start from fixed $\beta$ values along 
$\gamma = 1$ and determine the low-lying
spectrum as we systematically decrease $\beta_{5}$. The left panel of 
\fig{fig:b4_19_21} shows the approach
towards the bulk-driven phase transition for $\beta_{4} = 1.9$, where it is 
important to note that, even close to the phase transition,
we do not see any hint of finite volume effects. The black vertical line 
shows the location of the bulk-driven phase transition
separating the Higgs from the confined phase, while the orange horizontal 
line shows the Standard Model value for $\rho$.
It is clear that, in the vicinity of the phase transition, there is no 
qualitative change in the behaviour of $\rho$.

On the other-hand, the right panel of the same figure shows the same 
procedure but for $\beta_{4} = 2.1$. In this case, a major qualitative
change in the behaviour of the theory is apparent, as significantly, 
$\rho \sim 1.25$ close to the phase transition.
Note that in this case, in contrast to $\beta_4 = 1.9$, the phase transition 
separates the Higgs from the hybrid phase.
We find that on passing $\beta_{4} = 2.02$, which is the location of the 
boundary-driven phase transition, the physical properties of the theory
change such that the spectrum begins to resemble that of the Standard Model. 
We attribute this to the fact that beyond $\beta_{4} = 2.02$, the $U(1)$ 
gauge field on the boundary naturally exhibits a de-confined behaviour, 
as discussed in \sect{sec:phase_diagram}.

\begin{figure}[t]
  \begin{center}
   \includegraphics[width=1.0\textwidth]{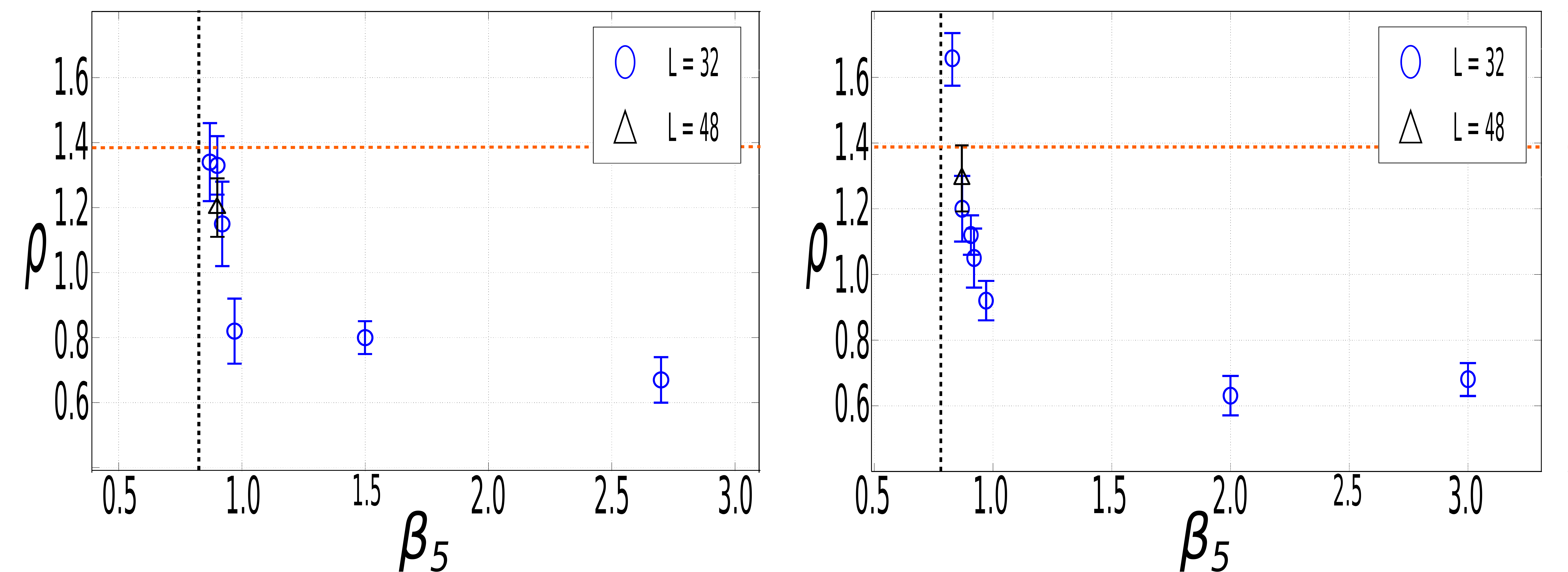}
  \end{center}
  \caption{\footnotesize The left (right) panel shows the ratio $\rho = m_{H}/ m_{Z}$ for various values of $\beta_{5}$
                         and fixed $\beta_{4} = 2.7$ ($3.0$). The calculations are performed on various volumes, $80 \times L^{3} \times 4$,
                         as described by the legend. The black vertical line shows the location of the Higgs-hybrid phase transition, while the orange horizontal line shows
                         the Standard Model value for $\rho$.}
  \label{fig:b4_27_30}
\end{figure}

\begin{figure}[t]
  \begin{center}
  \includegraphics[width=0.99\textwidth]{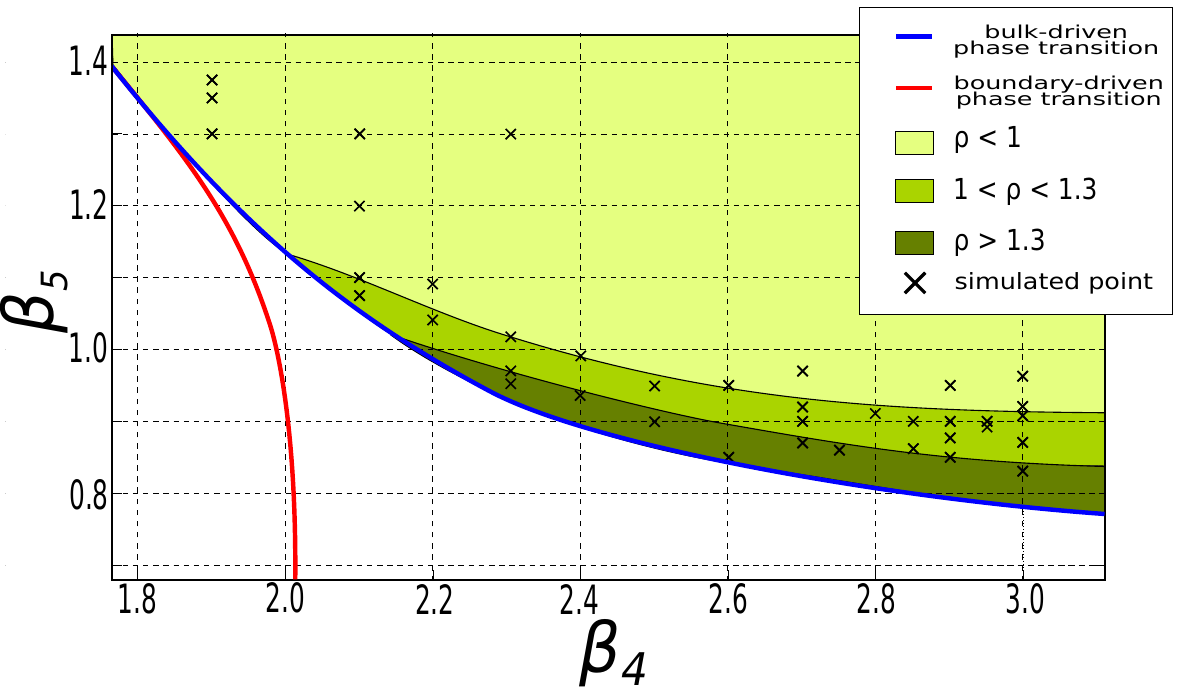}
  \end{center}
  \caption{\footnotesize Summary of our spectroscopic calculations for $\gamma < 1$ and $N_{5} = 4$ close to the Higgs-hybrid phase transition. The black points indicate
                         the location of a spectroscopic calculation within the phase diagram. The green shading indicates the $\rho \equiv m_{H}/m_{Z}$ value obtained for
                         a given calculation. The lightest shade corresponds to $\rho < 1$, the middle shade to $1 \leq \rho \leq 1.3$ and the darkest shade
                         corresponds to $\rho > 1.3$.}
  \label{fig:phase_diagram_with_rho}
\end{figure}

The left panel of \fig{fig:b4_27_30} shows the approach towards the 
phase transition for fixed $\beta_{4} = 2.7$. For $\beta_{5} = 0.9$,
we obtain a value of $\rho$ consistent with that of the Standard Model, $\rho = 1.33(9)$,
and in the right panel of the same figure we show that it is even possible for
the theory to produce values of $\rho$ significantly larger than the 
physical value of $1.38$. \fig{fig:phase_diagram_with_rho} shows a 
summary of our spectroscopic calculations close to the bulk-driven phase 
transition for $\gamma < 1$ and $N_{5} = 4$. As indicated by the legend, 
the different shadings of green correspond to three distinct regions in which 
we find $\rho < 1$ (lightest shade), $1 \leq \rho \leq 1.3$ (middle shade) and
$\rho > 1.3$ (darkest shade).

For various values of the couplings close to the bulk-driven phase transition, 
we robustly determine an excited $Z'$ state. Within this region, we find that
the $Z'$ is roughly three times heavier than the $Z$. However, we see evidence 
that, as we lower $\gamma$, the separation between $Z$ and $Z'$ increases. This
suggests that, in order to determine a more phenomenologically viable mass for 
$Z'$, we must continue our exploration close to the Higgs-hybrid phase 
transition to larger values of $\beta_{4}$. We note that, although we do not 
robustly determine an excited Higgs state within our explored range of 
couplings, we see evidence that close to the Higgs-hybrid phase transition it 
has a mass around four times that of the Higgs.

\section{First Steps Towards Varying the Size of the Extra Dimension}\label{sec:changing_n5}

As in all extra-dimensional models, it is crucial to determine how the physical properties of the theory
depend on the size of the extra dimension. In this section, we take first steps towards answering this question
by varying the parameter $N_{5}$ for a set of fixed couplings. Due to their qualitative differences, we again discuss the two regions $\gamma = 1$ and
$\gamma < 1$ separately.

\subsubsection*{\center{\bm{$\underline{\gamma = 1}$}}}\label{sec:N5_isotropic}

\begin{figure}[t!]
  \begin{center}
   \includegraphics[width=0.9\textwidth]{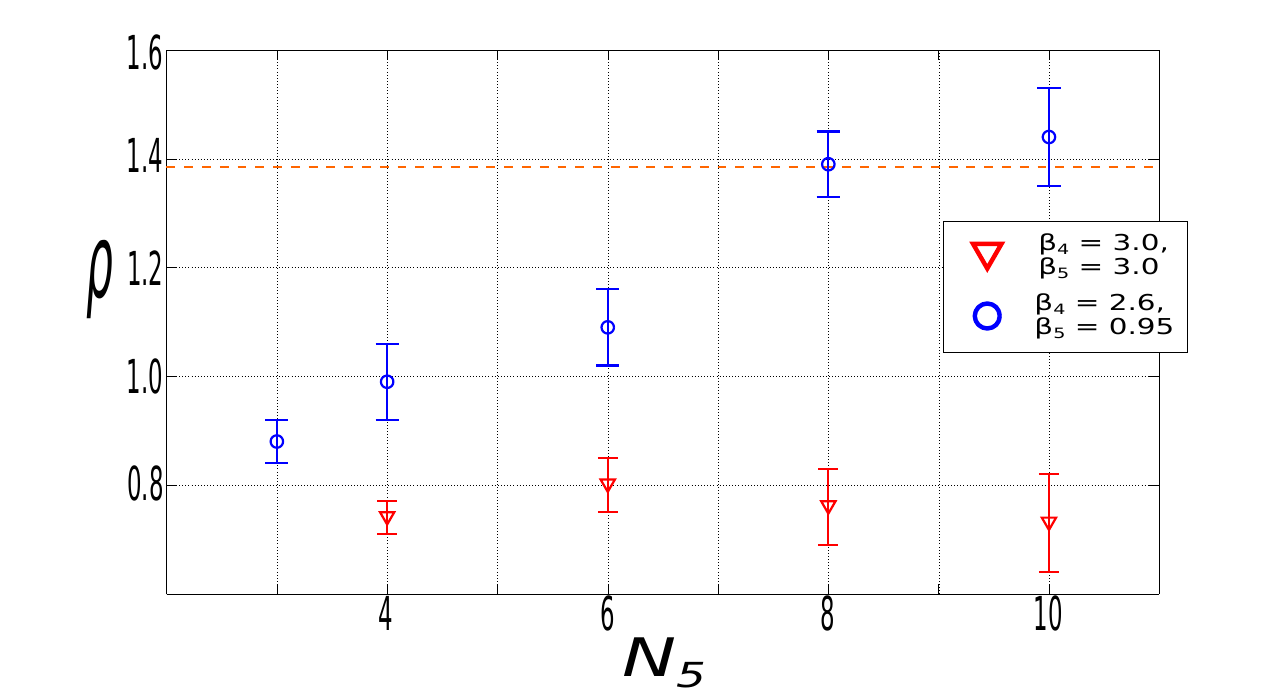}
  \end{center}
  \caption{\footnotesize The ratio $\rho$ for $\beta = 3.0$ ($\gamma = 1$, red triangles) and $\beta_{4} = 2.6$, $\beta_{5} = 0.95$ ($\gamma \approx 0.6$, blue circles) for various values of the extent of the extra dimension $N_{5}$.
                         The orange line corresponds to the Standard Model value of $\rho \approx 1.38$. The values were computed using $5,000$ gauge field configurations separated by $5$ HOR steps.}
   \label{fig:spectrum_vs_N5}
\end{figure}

The red triangles in \fig{fig:spectrum_vs_N5} shows the dependence of the ratio $\rho$ on the parameter $N_{5}$
for fixed isotropic coupling $\beta = 3.0$. As shown in \tab{tab:N5_scan_3.0}, both the Higgs and $Z$ boson masses remain relatively constant in lattice units, resulting in the constant behaviour of $\rho$ shown in the figure. We observe this behaviour for the range of isotropic couplings we explore and hence, it appears seemingly unlikely that the theory will produce a Standard
Model-like $\rho$ for any $N_{5}$ while $\gamma = 1$.

\subsubsection*{\center{\bm{$\underline{\gamma < 1}$}}}\label{sec:N5_low_gamma}

We now turn our attention to the $\gamma < 1$ regime, where we have observed that a Standard
Model-like hierarchy is possible close to the Higgs-hybrid phase transition. We note that the location of this transition slightly moves into the Higgs phase when $N_{5}$ changes from four to six.
Therefore, for the study of the $N_{5}$ dependence, we choose the couplings $\beta_{4} = 2.6$ and $\beta_{5} = 0.95$ as they are deep enough into the Higgs phase to avoid a change of phase as $N_{5}$ is increased. However, it should be noted that this shift seems to become negligible for higher $N_{5}$ values. The blue circles in \fig{fig:spectrum_vs_N5} show the dependence of the ratio $\rho$ on the extent of the extra dimension $N_{5}$. \tab{tab:N5_scan_2.6} shows that for increasing $N_{5}$, the Higgs mass slowly decreases in lattice units. However, the mass of the $Z$ boson rapidly decreases below the Higgs producing a Standard Model hierarchy for $N_{5} = 8, 10$. This feature of the theory may become crucial in future work, where we attempt to construct lines of constant physics with a Standard Model hierarchy, as it opens up a much larger parameter space. 

\begin{figure}[t!]
  \begin{center}
   \includegraphics[width=0.9\textwidth]{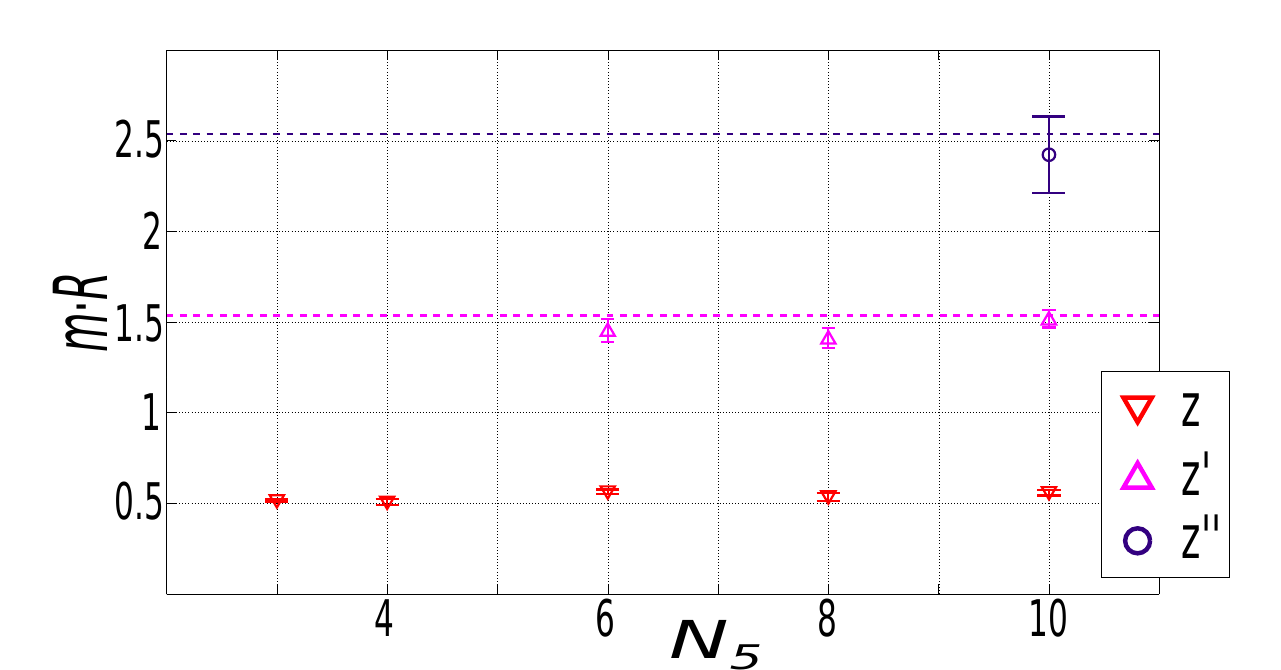}
  \end{center}
  \caption{\footnotesize The masses, in units of $1/R$, of the $Z$ (red triangles), $Z'$ (magenta triangles) and $Z''$ (purple circles) determined for a range of $N_{5}$
                         values at $\beta_{4} = 2.6$ and $\beta_{5} = 0.95$. The magenta (purple) dashed line corresponds to the mass expected
                         for the $Z'$ ($Z''$) from a Kaluza-Klein scenario. }
  \label{fig:k-k_vs_N5}
\end{figure}

An interesting observation is that the $Z$ channel exhibits a Kaluza-Klein-like behaviour. \fig{fig:k-k_vs_N5} shows the masses of the $Z$, $Z'$ and $Z''$ (where they could be robustly determined) for a range of $N_{5}$ values at $\beta_{4} = 2.6$ and $\beta_{5} = 0.95$. As shown in \tab{tab:N5_scan_2.6}, the masses $\sim 1/N_{5}$ in lattice units, producing the constant behaviour shown in the figure. Furthermore, we observe that the masses of the $Z'$ and $Z''$ fit the excitation tower, $m^{KK}_{i}$, expected from a Kaluza-Klein scenario. That is, that excitations $m^{KK}_{i+1}$ appear at energy scales $1/R$ higher than $m^{KK}_{i}$. Although we see hints of an excited Higgs state which lies around $1/R$ above the ground state, we do not robustly determine its mass and hence do not show it in \fig{fig:k-k_vs_N5}. Work is
under way in order to address this issue.

\section{Summary and Outlook} \label{sec:summary}

We have explored the phase diagram and computed the spectrum of a five-dimensional pure $SU(2)$ gauge theory formulated on a lattice orbifold
using anisotropic gauge couplings.
 
In terms of the phase diagram, our results are summarised by \fig{fig:phase_diag}. We find a first-order bulk-driven phase transition
separating confined and de-confined phases. This line is analogous to the one present within a toroidal geometry. 
However, due to the existence of an extra global symmetry within the orbifold, a non-perturbative SSB mechanism is present, in accordance with Elitzur's theorem.
The de-confined phase therefore becomes a Higgs phase due to the presence of massive gauge bosons.

Since the orbifold breaks explicitly the gauge symmetry from $SU(2)$ to $U(1)$ at the boundaries, 
we find the appearance of a boundary-driven phase transition which separates the confined from a third phase of the theory.
We label this phase as \textit{hybrid} due to its properties; the boundary layers behave like a four-dimensional $U(1)$ theory
in the Coulomb phase, while the four-dimensional bulk layers behave like a confining $SU(2)$ theory. This phase exhibits a layered-like
behaviour where each layer is (almost) independent of any other, and the physical content of a given layer is governed only by the gauge group of that
layer. \\

We have computed the spectrum of scalar (Higgs) and vector ($Z$) bosons in the isotropic $\gamma = 1$ and anisotropic $\gamma < 1$ regions. 
In the isotropic case, we find that the mass of the $Z$ boson is always larger than that of the Higgs. However, in the anisotropic region
close to the Higgs-hybrid phase transition, we find a Standard Model-like hierarchy of the Higgs and $Z$ masses. This appears only 
to be possible once the boundary gauge field \textit{naturally} de-confines, that is, once the four-dimensional coupling on the boundaries is larger than the critical
value associated with the four-dimensional $U(1)$ phase transition.

By computing the potential between two static charges in layers orthogonal to the extra dimension, we find that the system undergoes dimensional reduction precisely
in the region close to the Higgs-hybrid phase transition where we find the Standard Model-like hierarchy; the four-dimensional Standard Model-like physics is
\textit{localised} on the boundaries, while the bulk remains five-dimensional. In the bulk, we find the potential to be of a $5$-d Yukawa type, and the associated Yukawa mass
to agree with the $Z$ boson mass measured from our spectroscopic calculations. Significantly, the potential on the boundaries exhibits a $4$-d Yukawa form, where the associated 
Yukawa mass again matches that of the $Z$ boson. \\

The next steps in this project will address the issue of non-renormalisability of five-dimensional gauge theories.
In the phase diagram we only see first-order phase transitions. On the other hand, we seem to need a finite cut-off for spontaneous symmetry breaking
to occur. We will therefore have to establish whether lines of constant physics (LCPs) exist for a finite range of lattice spacings. Furthermore, these LCPs can
be compared to those of the four-dimensional Abelian Higgs model, which is the model that this theory is expected to reduce to in four-dimensions.

As a first step, we presented the dependence of the masses on the size of the extra dimension. The results appear favourable, and it seems that changing the
size of the extra dimension will open up a much larger Standard Model-like parameter space for constructing LCPs. One requirement will be to determine the physical
anisotropy, which is required to compute the physical size of the extra dimension.

Work in the direction of the LCPs at $\gamma < 1$, along with a study of the $\gamma > 1$ regime, in which the theory is expected to compactify, is already under-way.

\bigskip

{\bf Acknowledgements.}
We thank Peter Dziennik and Kyoko Yoneyama for their contributions
in an early stage of this work.
This work was funded by the Deutsche Forschungsgemeinschaft (DFG) under
contract KN 947/1-2. In particular G. M. acknowledges full support from the DFG.
The research of N.I. is implemented under the ARISTEIA II action
of the operational programme education and long life learning and
is co-funded by the European Union (European Social Fund) and National Resources.
The Monte Carlo simulations were carried out on
CHEOPS, a scientific supercomputer sponsored by the DFG
of the regional computing centre of the University of Cologne (RRZK)
and on the cluster Stromboli at the University of Wuppertal.
We thank both Universities for their support.
We are grateful to the  Astroparticle Physics group of the University of
Wuppertal for the use their graphics cards.
Finally we thank the ALPHA collaboration for providing software to compute
statistical errors of derived observables.

\begin{appendix}
\section{Comments on the Stick Symmetry}
\label{app_stick}

The stick symmetry ${\cal S}_{L}$ is defined by the transformations
\eqref{eq:st1} and \eqref{eq:st2}.
Without loss of generality we can take $g_s=-i \sigma^1$.
Notice that while \eq{eq:st2} is a $\mathbb{Z}_2$ transformation, \eq{eq:st1} is a
$\mathbb{Z}_4$ transformation.
In particular the order 2 element of this $\mathbb{Z}_4$ takes a hybrid link into
minus itself, yielding a transformation by a centre element.
We denote the centre of SU(2) by $\mathbb{Z}_2^c$.
The scalar and vector Polyakov loops defined in \sect{sec:operators}
are invariant under this $\mathbb{Z}_2^c$ (unlike on the periodic lattice).
This specific part of the centre symmetry does not have any physical
consequences, either of the finite temperature class or any other.
Therefore the nontrivial part of the
symmetry is obtained by
taking the quotient $\mathbb{Z}_4/\mathbb{Z}_2^c$ which is isomorphic to $\mathbb{Z}_2$.
Thus, the stick symmetry is indeed a $\mathbb{Z}_2$ symmetry, identified as the
generalized Weyl group \cite{Irges:2013rya}.
A similar discussion applies to ${\cal S}_R$.

We can attempt to follow the fate of the stick symmetry in the continuum limit.
Assuming that the only continuum limit is the perturbative limit, we can write
a hybrid link as $U(n_5=0,5) = e^{i a_5(\sigma^1 A_5^1 +\sigma^2 A_5^2 )}$ where
$a_5$ is the lattice spacing along the fifth dimension.
Then, defining $\phi = i A_5^1 + A_5^2$ and expanding in the lattice spacing and
keeping only the leading order terms, we obtain from \eq{eq:st1}

\begin{align}
1 \longrightarrow ia_5\phi\,\,\,\,\,\,\,     &&      ia_5\phi   \longrightarrow -1 \nonumber\\
1 \longrightarrow -ia_5\phi^*  &&    -ia_5\phi^*  \longrightarrow -1\label{eq:contst}
\end{align}

These transformations are consistent if $A_5^2=0$, that is if $\phi = i A_5^1$.
Had we chosen for $g_s$ the most general linear combination
$g_s = -i (\cos\theta\, \sigma^1 + \sin\theta\, \sigma^2)$,
we would have obtained the transformations corresponding to the
linear combination $\phi = (i\cos\theta+\sin\theta)A$,
where $A$ is a real field.

The transformation of the boundary links in \eq{eq:st2} looks like a global
$U(1)$ transformation so it has a well defined continuum limit.
The two conclusions we can draw from the transformation of the hybrid links in
\eq{eq:contst} are that
they are non-trivial only because the fluctuating degrees of freedom are
group elements
and that they interchange a field by a constant. The latter is consistent
with the field taking a vacuum expectation value.
The first conclusion on the other hand implies that in the naive, perturbative
continuum limit one expects the stick symmetry together with SSB to disappear.
Hence, the Higgs mechanism on the (pure gauge) orbifold lattice in the
interior of the phase diagram seems to be tied to
a non-zero lattice spacing which is in fact consistent with the presence of
the first-order, bulk phase transition.
The only way to circumvent this is if a second-order phase transition is found,
where a non-trivial continuum limit could be taken.
In such a case however one would have to possibly change the expansion
of the link in terms of the continuum fields and the analysis would have to
be repeated.
Since at present we do not see any second-order phase transition, we will
not discuss further this possibility.

\section{Location of Phase Transitions \label{app:phase_transition}}
In this appendix we give the location of the theory's phase transitions (see \tab{tab:bulkpt}) determined on $L^4\times4$ lattices.
The listed values refer to the centre of the hysteresis, observed from the expectation value of the plaquette  after \textit{hot}
and \textit{cold} starts. The uncertainty associated with the values is given by half the observed width of the hysteresis.
The plaquette expectation values were measured on $1000$ gauge field configurations, separated by $2$ HOR steps.
Note that volume checks have been performed and the phase transitions have been found to not
show any dependence on $L$ once $L \geq 32$.


\begin{table}[b!]
\begin{centering}
\begin{minipage}{0.45\textwidth}
\begin{tabular}{@{\extracolsep{0.2cm}}ccc}
\toprule
\multicolumn{3}{c}{Bulk-driven PT} \\
\midrule
$L$ & $\beta_4$ & $\beta_5$ \\
\midrule
32   &   1.45  & 1.95(10) \\
32   &   1.5   & 1.85(10) \\
32   &   1.615(45)   & 1.615(45) \\
32   &   1.70  & 1.48(8)             \\
32   &   1.80  & 1.35(5)             \\
\rowcolor{lightgray} 32   &   1.90  & 1.21(3)             \\
32   &   1.92  & 1.20(1)             \\
32   &   1.93  & 1.19(1)             \\
32   &   1.94  & 1.188(13)           \\
32   &   1.95  & 1.17(1)             \\
32   &   1.96  & 1.1625(75)          \\
32   &   1.97  & 1.155(15)           \\
32   &   1.98  & 1.15(1)             \\
32   &   1.99  & 1.14(1)             \\
32   &   2.0  & 1.13(1)             \\
32   &   2.01  & 1.125(15)           \\
32   &   2.02  & 1.11(1)             \\
32   &   2.03  & 1.10(1)             \\
32   &   2.04  & 1.10(1)             \\
32   &   2.06  & 1.08(1)             \\
32   &   2.08  & 1.065(5)            \\
40   &   2.10  & 1.05(1)             \\
40   &   2.15  & 1.015(5)            \\
40   &   2.20  & 0.985(5)            \\
40   &   2.25  & 0.955(5)            \\
40   &   2.30  & 0.93(1)             \\
40   &   2.40  & 0.890(5)            \\
\bottomrule
\end{tabular}

\end{minipage}
\hfill
\begin{minipage}{0.45\textwidth}
\begin{tabular}{@{\extracolsep{0.2cm}}ccc}
\toprule
\multicolumn{3}{c}{Bulk-driven PT} \\
\midrule
$L$ & $\beta_4$ & $\beta_5$ \\
\midrule{}
40   &   2.50  & 0.8625(75)          \\
40   &   2.55  & 0.8475(25)          \\
40   &   2.6  & 0.8387(38)          \\
40   &   2.7  & 0.8200(50)          \\
40   &   2.8  & 0.8050(50)          \\
40   &   2.9  & 0.7900(50)          \\
40   &   3.0  & 0.7800(50)          \\
\bottomrule
\end{tabular}

\vspace{0.8cm}

\begin{tabular}{@{\extracolsep{0.2cm}}ccc}
\toprule
\multicolumn{3}{c}{Boundary-driven PT} \\
\midrule
$L$ & $\beta_4$ & $\beta_5$ \\
\midrule
\rowcolor{lightgray} 32   &   1.9    &   1.21(3) \\
32   &   1.92   &   1.175(15) \\
32   &   1.93   &   1.155(15) \\
32   &   1.94   &   1.137(13) \\
32   &   1.95   &   1.12(1)   \\
32   &   1.96   &   1.09(1)   \\
32   &   1.97   &   1.065(15) \\
32   &   1.98   &   1.035(15) \\
32   &   1.99   &   0.99(2)   \\
32   &   2.0    &   0.925(25) \\
32   &   2.01   &   0.825(75) \\
32   &   2.02(1)   &  0.75    \\
32   &   2.02(1)   &  0.5     \\
32   &   2.02(1)   &  0.2     \\
32   &   2.02(1)   &  0.01    \\
\bottomrule
\end{tabular}

\end{minipage}




\caption{\label{tab:bulkpt}\footnotesize The location of the
first-order phase transitions for the $N_5=4$ orbifold.
Where indicated, the errors correspond to half of the size of the hysteresis.
The triple-point is highlighted by the the light grey background.}

\end{centering}
\end{table}

\clearpage

\section{Tables of Masses \label{app:eff_masses}}

In this appendix we tabulate, in lattice units, the masses obtained by the simulations described in Sections \ref{sec:spectrum} and \ref{sec:changing_n5}.
\tab{tab:masses} lists the masses measured on lattices where the extent of the extra dimension $N_5 = 4$, for different values of the couplings $\beta_4$ and $\beta_5$.
Tables \ref{tab:N5_scan_3.0}  and \ref{tab:N5_scan_2.6} show the dependence on $N_5$ of masses measured at $\beta = 3.0,~\gamma=1$, and $\beta_4 = 2.6,~\beta_5 = 0.95$ respectively.

\begin{longtable}[]{cccccccc}
\toprule

$\beta_4$ & $\beta_5$ & Lattice volume &   $a_{4}m_H$     &   $a_{4}m_{Z}$     &    $\rho$      & $a_{4}m_{Z'}$ \\
\midrule
\endhead
\bottomrule
\endfoot
\kill
\endlastfoot
1.66   &  1.66   &  $80\text{x}32^{3}\text{x}4$    &  0.217(35)    &  0.466(34)  &  0.59(8)      & -             \\
1.66   &  1.66   &  $80\text{x}48^{3}\text{x}4$    &  0.246(16)    &  0.418(18)  &  0.59(8)      & -             \\
1.7    &  1.7    &  $80\text{x}32^{3}\text{x}4$    &  0.235(31)    &  0.423(27)  &  0.55(9)      & -             \\
1.75   &  1.75   &  $80\text{x}32^{3}\text{x}4$    &  0.241(31)    &  0.343(56)  &  0.70(12)     & -             \\
1.8    &  1.8    &  $80\text{x}32^{3}\text{x}4$    &  0.199(10)    &  0.317(14)  &  0.61(5)      & -             \\
1.9    &  1.3    &  $64\text{x}48^{3}\text{x}4$    &  0.204(25)    &  0.433(5)   &  0.47(6)      & -             \\
1.9    &  1.3    &  $80\text{x}32^{3}\text{x}4$    &  0.195(6)     &  0.422(5)   &  0.46(8)      & -             \\
1.9    &  1.6    &  $80\text{x}32^{3}\text{x}4$    &  0.239(26)    &  0.361(34)  &  0.66(15)     & -             \\
1.9    &  1.9    &  $80\text{x}32^{3}\text{x}4$    &  0.245(26)    &  0.358(24)  &  0.68(8)      & -             \\
2.0    &  2.0    &  $80\text{x}48^{3}\text{x}4$    &  0.245(21)    &  0.403(25)  &  0.61(8)      & -             \\
2.1    &  1.075  &  $80\text{x}32^{3}\text{x}4$    &  0.327(8)     &  0.268(3)   &  1.22(4)      & -             \\
2.1    &  1.1    &  $80\text{x}32^{3}\text{x}4$    &  0.223(7)     &  0.181(10)  &  1.23(8)      & -             \\
2.1    &  1.1    &  $80\text{x}48^{3}\text{x}4$    &   0.215(19)   &  0.183(21)  &  1.19(12)     & -             \\
2.1    &  1.2    &  $80\text{x}32^{3}\text{x}4$    &  0.192(18)    &  0.351(14)  &  0.54(6)      & -             \\
2.1    &  1.3    &  $80\text{x}32^{3}\text{x}4$    &  0.243(7)     &  0.405(23)  &  0.60(5)      & -             \\
2.1    &  1.3    &  $64\text{x}48^{3}\text{x}4$    &  0.239(21)    &  0.415(25)  &  0.58(8)      & -             \\
2.1    &  1.6    &  $80\text{x}32^{3}\text{x}4$    &  0.237(22)    &  0.337(23)  &  0.71(8)      & 0.550(12)     \\
2.2    &  1.04   &  $80\text{x}32^{3}\text{x}4$    &  0.231(14)    &  0.242(19)  &  0.96(8)      & -             \\
2.2    &  1.09   &  $80\text{x}32^{3}\text{x}4$    &  0.220(11)    &  0.299(24)  &  0.75(7)      & 0.49(5)       \\
2.3    &  0.95   &  $80\text{x}32^{3}\text{x}4$    &  0.215(5)     &  0.187(2)   &  1.15(3)      & -             \\
2.3    &  0.97   &  $80\text{x}32^{3}\text{x}4$    &  0.229(10)    &  0.217(15)  &  1.06(6)      & -             \\
2.3    &  1.02   &  $80\text{x}32^{3}\text{x}4$    &  0.239(11)    &  0.236(12)  &  1.01(8)      & 0.525(28)     \\
2.3    &  1.3    &  $80\text{x}32^{3}\text{x}4$    &  0.277(26)    &  0.386(22)  &  0.71(4)      & -             \\
2.4    &  0.94   &  $80\text{x}32^{3}\text{x}4$    &  0.214(17)    &  0.204(8)   &  1.05(9)      & -             \\
2.4    &  0.99   &  $80\text{x}32^{3}\text{x}4$    &  0.204(13)    &  0.250(12)  &  0.82(6)      & -             \\
2.5    &  0.9    &  $80\text{x}32^{3}\text{x}4$    &  0.197(8)     &  0.184(5)   &  1.07(6)      & -             \\
2.5    &  0.95   &  $80\text{x}32^{3}\text{x}4$    &  0.223(12)    &  0.222(18)  &  1.01(9)      & 0.49(3)       \\
2.6    &  0.85   &  $80\text{x}32^{3}\text{x}4$    &  0.208(13)    &  0.128(3)   &  1.67(11)     & 0.351(17)     \\
2.6    &  0.95   &  $80\text{x}32^{3}\text{x}4$    &  0.238(9)     &  0.240(8)   &  0.99(7)      & 0.58(5)       \\
2.7    &  0.87   &  $80\text{x}32^{3}\text{x}4$    &  0.233(23)    &  0.175(8)   &  1.34(12)     & 0.546(25)     \\
2.7    &  0.9    &  $80\text{x}32^{3}\text{x}4$    &  0.267(15)    &  0.201(8)   &  1.33(9)      & 0.37(5)       \\
2.7    &  0.9    &  $80\text{x}48^{3}\text{x}4$    &  0.250(10)    &  0.208(16)  &  1.20(9)      & 0.550(17)     \\
2.7    &  0.92   &  $80\text{x}32^{3}\text{x}4$    &  0.26(3)      &  0.22(1)    &  1.15(13)     & -             \\
2.7    &  0.97   &  $80\text{x}32^{3}\text{x}4$    &  0.19(2)      &  0.23(1)    &  0.82(10)     & -             \\
2.7    &  1.5    &  $80\text{x}32^{3}\text{x}4$    &   0.221(7)    &  0.274(15)  &  0.80(5)      & 0.424(5)      \\
2.7    &  2.7    &  $80\text{x}32^{3}\text{x}4$    &  0.212(18)    &  0.316(12)  &  0.67(7)      & -             \\
2.75   &  0.86   &  $80\text{x}32^{3}\text{x}4$    &  0.207(11)    &  0.174(6)   &  1.19(8)      & 0.452(35)     \\
2.8    &  0.91   &  $80\text{x}32^{3}\text{x}4$    &  0.24(5)      &  0.20(2)    &  1.18(16)     & -             \\
2.8    &  2.8    &  $80\text{x}32^{3}\text{x}4$    &  0.236(12)    &  0.356(18)  &  0.66(5)      & -             \\
2.85   &  0.862  &  $80\text{x}32^{3}\text{x}4$    &  0.222(13)    &  0.179(8)   &  1.24(9)      & -             \\
2.85   &  0.9    &  $80\text{x}32^{3}\text{x}4$    &   0.216(11)   &  0.207(8)   &  1.04(6)      & -             \\
2.9    &  0.85   &  $80\text{x}32^{3}\text{x}4$    &  0.200(10)    &  0.156(7)   &  1.28(9)      & -             \\
2.9    &  0.877  &  $80\text{x}32^{3}\text{x}4$    &  0.26(2)      &  0.216(4)   &  1.19(1)      & -             \\
2.9    &  0.9    &  $80\text{x}32^{3}\text{x}4$    &  0.28(2)      &  0.20(1)    &  1.36(10)     & -             \\
2.9    &  0.95   &  $80\text{x}32^{3}\text{x}4$    &  0.274(12)    &  0.256(8)   &  1.07(6)      & -             \\
2.95   &  0.892  &  $80\text{x}32^{3}\text{x}4$    &   0.238(13)   &  0.171(4)   &  1.39(8)      & -             \\
2.95   &  0.9    &  $80\text{x}32^{3}\text{x}4$    &  0.22(3)      &  0.21(2)    &  1.06(18)     & -             \\
3.0    &  0.83   &  $80\text{x}32^{3}\text{x}4$    &  0.263(11)    &  0.158(3)   &  1.66(8)      & 0.333(24)     \\
3.0    &  0.87   &  $80\text{x}32^{3}\text{x}4$    &  0.26(2)      &  0.216(9)   &  1.20(10)     & -             \\
3.0    &  0.87   &  $80\text{x}48^{3}\text{x}4$    &  0.249(14)    &  0.193(7)   &  1.29(8)      & 0.491(25)     \\
3.0    &  0.907  &  $80\text{x}32^{3}\text{x}4$    &  0.269(15)    &  0.241(6)   &  1.12(6)      & -             \\
3.0    &  0.92   &  $80\text{x}32^{3}\text{x}4$    &  0.253(23)    &  0.242(8)   &  1.05(9)      & 0.518(30)     \\
3.0    &  0.97   &  $80\text{x}32^{3}\text{x}4$    &  0.229(14)    &  0.253(10)  &  0.91(6)      & -             \\
3.0    &  2.0    &  $80\text{x}32^{3}\text{x}4$    &  0.203(14)    &  0.323(19)  &  0.63(6)      & 0.583(11)     \\
3.0    &  3.0    &  $80\text{x}32^{3}\text{x}4$    &  0.251(7)     &  0.335(11)  &   0.74(3)     & -             \\
4.0    &  4.0    &  $80\text{x}32^{3}\text{x}4$    &  0.233(7)     &  0.315(13)  &  0.74(4)      & -             \\
5.0    &  5.0    &  $80\text{x}48^{3}\text{x}4$    &  0.192(16)    &  0.297(15)  &  0.65(6)      & -             \\
\bottomrule
\caption{\label{tab:masses}\footnotesize{ Masses measured on $N_5 = 4$ lattices. } }

\end{longtable}

\begin{table}[t!]
\begin{center}
\begin{tabular}{@{\extracolsep{0.2cm}}cccc}
\toprule
$N_5$ & $ a_4 m_H$ & $a_4 m_Z$ & $ \rho $ \\
\midrule
4   &   0.251(7)    &    0.335(11)    &   0.74(3) \\
6   &   0.232(12)   &    0.289(13)    &   0.80(5) \\
8   &   0.205(12)   &    0.268(20)    &   0.76(7) \\
10  &   0.219(24)   &    0.297(11)    &   0.73(9) \\
\bottomrule
\end{tabular}
\caption{\label{tab:N5_scan_3.0}\footnotesize{ Masses determined at $\beta_4 = 3.0$, $\beta_5 = 3.0$ for $N_5 \in [4,10]$, on volumes $80 \times 32^{3} \times N_{5}$. }  }
\end{center}
\end{table}

\begin{table}[t]
\begin{center}
\begin{tabular}{@{\extracolsep{0.2cm}}cccccc}
\toprule
$N_5$ & $ a_4 m_H$ & $a_4 m_Z$ & $ \rho $ & $a_4 m_{Z'}$ & $a_4 m_{Z''}$\\
\midrule
3   &    0.284(12)    &   0.325(4)     &   0.88(4)  &  -          &    -   \\
4   &    0.238(9)     &   0.240(8)     &    0.99(7) &  0.58(5)    &    -   \\
6   &    0.193(11)    &   0.178(4)     &   1.09(7)  &  0.46(2)    &    -   \\
8   &    0.177(9)     &   0.127(5)     &   1.39(6)  &  0.335(13)  &    -   \\
10  &    0.152(8)     &   0.106(3)     &   1.44(9)  &  0.288(9)   &    0.46(4)   \\
\bottomrule
\end{tabular}
\caption{\label{tab:N5_scan_2.6}\footnotesize{ Masses determined at $\beta_4 = 2.6$, $\beta_5 = 0.95$ for $N_5 \in [3,10]$, on volumes $80 \times 32^{3} \times N_{5}$. } }
\end{center}
\end{table}

\vspace{2cm}

\end{appendix}

\bibliography{orbifold}
\bibliographystyle{JHEP}

\end{document}